\documentclass[11pt,twocolumn]{article}
\usepackage{graphicx}

\usepackage{lineno}
\usepackage{color}
\usepackage{cite}
\usepackage{multicol}
\makeatletter
\renewcommand{\@biblabel}[1]{\quad#1}
\makeatother

\usepackage{amssymb,amsfonts,amsmath}
\newcommand{\be}{\begin{eqnarray}}
\newcommand{\ee}{\end{eqnarray}}
\newcommand{\la}{\langle}
\newcommand{\ra}{\rangle}

\begin{document}


\title{Impact of epistasis and pleiotropy on evolutionary adaptation}

\author{Bj\o rn \O stman$^{1,2,3,\ast}$, Arend Hintze$^{1,3,4}$, Christoph Adami$^{1,2,3}$\\\mbox{}\\
$^1$ Keck Graduate Institute of Applied Life Sciences\\ Claremont, CA 91711\\
$^2$ Microbiology and Molecular Genetics\\
$^3$ BEACON Center for the Study of Evolution in Action\\
$^4$ Department of Computer Science and Engineering\\ Michigan State University, East Lansing, MI 48823\\
$\ast$ E-mail: ostman@msu.edu}
\onecolumn

\maketitle

\begin{abstract}
Evolutionary adaptation is often likened to climbing a hill or peak. While this process is simple for fitness landscapes where  mutations are independent, the interaction between mutations (epistasis) as well as mutations at loci that affect more than one trait (pleiotropy) are crucial in complex and realistic fitness landscapes. We investigate the impact of epistasis and pleiotropy on adaptive evolution by studying the evolution of a population of asexual haploid organisms (haplotypes) in a model of $N$ interacting loci, where each locus interacts with $K$ other loci. We use a quantitative measure of the magnitude of epistatic interactions between substitutions, and find that it is an increasing function of $K$.  When haplotypes adapt at high mutation rates, more epistatic pairs of substitutions are observed on the line of descent than expected. The highest fitness is attained in landscapes with an intermediate amount of ruggedness that balance the higher fitness potential of interacting genes with their concomitant decreased evolvability.  Our findings imply that the synergism between loci that interact epistatically is crucial for evolving genetic modules with high fitness, while too much ruggedness stalls the adaptive process.
\end{abstract}
\newpage
\twocolumn
\section*{Introduction}
As a population adapts to its environment, it accumulates mutations that increase the chance for the long-term success of the lineage (or lineages) it represents. The standard picture for this process is Fisher's geometric model~\cite{Fisher1930} of evolution by small steps, i.e., the accumulation of many mutations with small benefit. The evidence supporting this concept, however, is scarce~\cite{BurchChao1999}, and many open questions remain~\cite{Orr2005}. 

More modern treatments use stochastic substitution models~\cite{Gillespie1984,Gillespie1991,Orr2002,KimOrr2005,Kryazhimskiyetal2009} to understand the adaptation of DNA sequences. If the mutation rate is small and selection is strong, the adaptive process can explore at most a few mutational steps away from the wild type, so that mutations are fixed sequentially and deleterious mutations only plays a minor role (if any)~\cite{Kryazhimskiyetal2009}. However, if the rate of mutation is high (and/or selection is weak) mutations can interact significantly and adaptation  does {\em not} proceed solely via the accumulation of only beneficial (and neutral) mutations. Instead, deleterious mutations play an important role as stepping stones of adaptive evolution that allow a population to traverse fitness valleys. Kimura, for example, showed that a deleterious mutation can drift to fixation if followed by a compensatory mutation that restores fitness~\cite{Kimura1985}.
Recent work using computational simulations of evolution has shown that deleterious mutations are crucial for adaption, and interact with subsequent mutations to create substantial beneficial effects~\cite{Lenskietal2003,Bridgham2006,Poelwijk2006,Cowperthwaite2006,Clune2008}. Even though the potential of interacting mutations in adaptive evolution has been pointed out early by Zuckerkandl and Pauling~\cite{ZuckerkandlPauling1965}, their importance in shaping adaptive paths through a fitness landscape has only recently come to the forefront~\cite{BloomArnold2009,Phillips2008,WeinreichWatsonChao2005,Poelwijk2007}, and is still a topic of much discussion~\cite{Reetzetal2005,WeinreichChao2005,Weinreich2006,Lockless1999}. In this respect, the impact of the sign (i.e., positive or negative) as well as the {\em size} of epistasis on adaptation, and how this impact is modulated by the mutation rate,  has not received the attention it deserves~\cite{Whitlock1995,Coyneetal2000,Philippsetal2000}. 

If we move from the single gene level to networks of genes, the situation becomes even more complex. Gene networks that have been explored experimentally are strongly epistatic~\cite{KelleyIdeker2005,UlitskyShamir2007,Roguevetal2008,Costanzoetal2010}, and allelic changes at one locus significantly modulate the fitness effect of a mutation at another locus. To understand the evolution of such systems, we have to take into account the interaction between loci, and furthermore abandon the limit where mutations on different loci fix sequentially. Here, we quantify the impact of epistasis on evolutionary adaptation (and the dependence of this impact on mutation rate),  by studying a computational model of a fitness landscape of $N$ loci, whose ruggedness can be tuned: the NK landscape model of Kauffman~\cite{KauffmanLevin1987,KauffmanWeinberger1989,Kauffman1993,Altenberg1997}.  The model (and versions of it known as the ``blocks model") has been used to study a variety of problems in evolution (see, e.g.,~\cite{KauffmanWeinberger1989,MackenPerelson1989,PerelsonMacken1995,Solowetal1999,Campos2002,WelchWaxman2005,Orr2006}), but most concern the evolution of beneficial alleles at a single locus. Models that study interacting gene networks (for example, transcriptional regulatory networks) have focused mainly on the topology, robustness, and modularity of the network~\cite{Wagner1996,Cilibertietal2007,Azevedoetal2006,EspinosaSoto2010}. Instead, we are interested in the evolution of the allelic states of the network as a population evolves from low fitness to high fitness: how interacting mutations allow the crossing of fitness valleys, and how the ruggedness of the landscape shapes the evolutionary path.  The NK model can describe genetic interactions that are more complex than what can be achieved within standard population genetics in which no more than two-locus models are tractable (here, we use $N=20$ loci), while we retain the ability to carry out simulations with high statistics.

As opposed to most work studying adaptation in the NK fitness landscape, we do not focus on population observables such as mean fitness, but rather study the line of descent in each population in order to characterize the sequence and distribution of mutations that have come to represent the evolutionary path (see, e.g., \cite{Lenskietal2003,Cowperthwaite2006}). We consider this approach more valuable because it more closely mimics studies in nature where usually the information we gain about about evolutionary history is from surviving lineages. The mutations that are found on the line of descent are not independent of each other in general, and paint a complex picture of adaptation that involves deleterious and beneficial mutations that are conditional on the presence of each other and other alleles on the haplotype, of valley crossings, compensatory mutations, and reversals. Further, we are specifically interested in evolution in landscapes of intermediate ruggedness, since fitness of neighboring genotypes in maximally rugged landscapes of $K=N-1$ are uncorrelated (see~\cite{JainKrug2007} for an analysis of evolution in maximally rugged landscapes).

\subsection*{NK model}
The NK model of genetic interactions~\cite{KauffmanLevin1987,KauffmanWeinberger1989,Kauffman1993} consists of circular, binary sequences encoding the alleles at $N$ loci, where each locus contributes to the fitness of the haplotype via an interaction with $K$ other loci. For each of the $N$ loci, we create a lookup-table with random numbers between 0 and 1 (drawn from a uniform distribution) that represent the fitness contribution $w_i$ of a binary sequence of length $K+1$. For example, the case $K=1$ (interaction with one other locus) is modeled by creating random numbers for the four possible binary pairs {\tt 00,01,10,11} for each of the $N$ loci, that is, the fitness contribution at one locus is conditional on the allele at one other locus (usually adjacent). Because independent random numbers are drawn for the four different combinations, the fitness contribution of a locus to the overall fitness of the organism can change drastically depending on the allele of the interacting locus. The case $K=0$  is the simplest (no interacting loci apart form reversals, and therefore vanishing epistasis). This choice gives rise to a smooth landscape with only a single peak that any search algorithm can locate in linear time, whereas increasing $K$ makes the fitness of a locus dependent on a total of $K+1$ loci, resulting in a rugged landscape with multiple local peaks. At the same time, the fitness of $K+1$ loci is affected by a single mutation, giving rise to pervasive pleiotropy that amplifies the ruggedness of the landscape by increasing the effect of single mutations. Pleiotropy is one of the main assumptions behind Fisher's geometric model, and appears to be common in nature~\cite{Ostrowski2005,Wagneretal2008}. Increasing $K$ increases ruggedness, i.e., it increases both the number of peaks (frequency) and the variation in genotype fitness (amplitude). In the NK model, the increased peak amplitude is caused by  pleiotropy: when each fitness component is determined by $K+1$ loci, then it is also true that each locus acts pleiotropically, affecting $K+1$ fitness components. Since the lookup-tables contain $2^{K+1}$ random numbers, the likelihood of finding a haplotype of very high fitness increases with $K$. The average height of the global peak thus increases with $K$~\cite{Skellettetal2005}. For this reason we expect that adaptation will result in higher fitness when loci interact more, as long as the evolutionary dynamics allows the population to locate the higher peaks. We argue that this effect is not solely an artifact of this model, but that it is also an effect that should be observable empirically (see Discussion). However, we also provide control simulations where the height of the fitness peaks is normalized. 

\begin{figure}[htbp]
\centering
\includegraphics[width=0.5\textwidth]{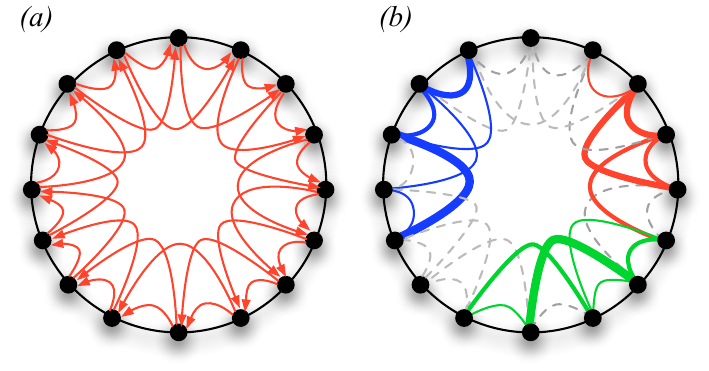}
\caption{NK model haplotypes for $N=16$ and $K=2$. For these parameters, the fitness contribution of each locus is determined by interacting with 2 loci (adjacent in the representation shown here), giving rise to blocks of $2K+1$ interacting genes.  {\em (a)}:  Interactions between loci represented by lines. {\em (b)}: Actual epistatic interactions on a particular high-fitness peak, where the width of the lines indicates the strength of epistatic interactions (thicker lines equal higher values of $\varepsilon$, defined below). Three modules of interacting loci are coloured. The remaining interactions (dashed grey lines) are weak.}
\label{NK-ring}
\end{figure}

In Fig.~\ref{NK-ring}a, we show an example haplotype with $N=16$ and $K=2$, indicating the potential interactions. For high-fitness haplotypes, some interactions are stronger than others (Fig.~\ref{NK-ring}b), and lead to the formation of clusters of strongly interacting loci (modules).

While the NK model is an abstract model of a fitness landscape, the number of interacting genes that we consider ($N=20$) is comparable to viruses (e.g., HIV has 15 proteins, see~\cite{FrankelYoung1998}), or else to modular pathways whose function directly affects the fitness of the organism. Genetic networks with modular structure are common in living organisms~\cite{Hanetal2004}, and examples of modules with approximately 20 genes or proteins include fibrin blood clotting with 26 genes~\cite{Doolittle2009} and human mitochondria with 37 genes~\cite{Andersonetal1981}. The modular composition of such structures ensures that selection can act on them without affecting other traits at the same time, and the breaking of pleiotropic constraints between modules coding for separate traits is thought to result in networks with a high level of modular partitioning~\cite{WagnerAltenberg1996}.

Here we choose the fitness of each haplotype to be the {\em geometric mean} of the values $w_i$ found in the lookup-tables, 
\be
W = \left(\prod_{i=1}^{N}w_i\right)^{1/N}, \label{land}
\ee
rather than the {\em average} as is done traditionally~\cite{Kauffman1993}.  This form is more realistic than its additive counterpart and has been suggested before~\cite{Solowetal2000,WelchWaxman2005}. In such a landscape, single mutations can potentially have a large effect on fitness, including lethality. The landscape defined by Eq.~(\ref{land}) gives rise to very few neutral mutations because each locus contributes to fitness in one way or the other, and we have not explicitly introduced alleles with zero fitness (lethals). We note that the results presented here do not depend on whether fitness is the arithmetic or the geometric mean (electronic supplementary material figure S1). 

\subsection*{Quantifying Epistasis}
\begin{figure*}[!htbp]
\center\includegraphics[width=0.8\textwidth]{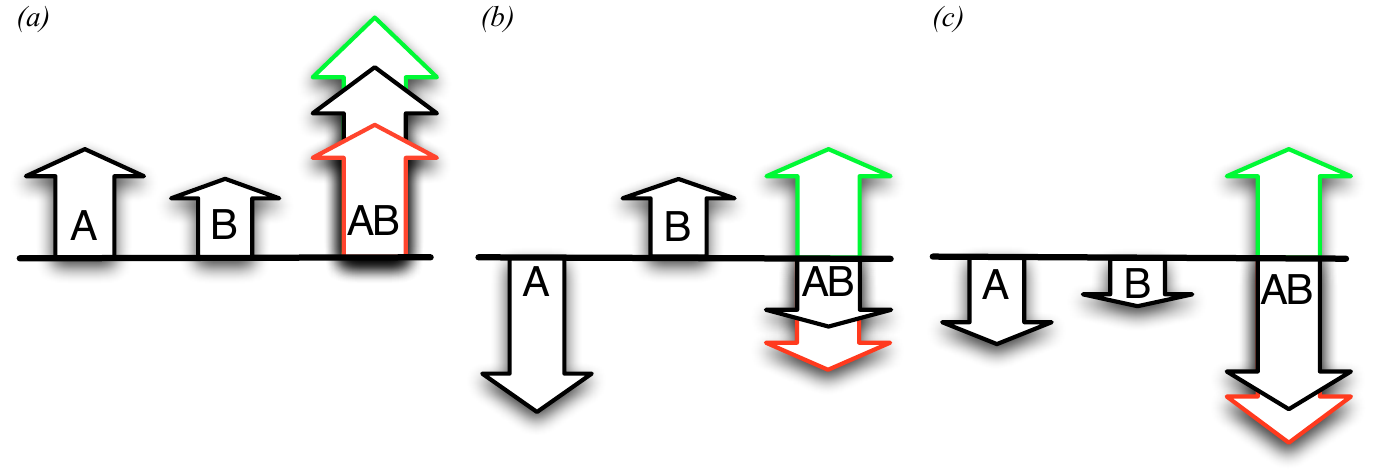}
\caption{Schematic representation of epistasis. Two mutations A and B can interact epistatically in different ways with varying effects on fitness. The fitness of the wild-type is represented by the black baselines, and the heights of arrows represent the fitness after one mutation ($ W_A$ or $ W_B$) and after both mutations ($ W_{AB}$). Green: positive epistasis,  red: negative epistasis. In {\em (a)}, two independently beneficial mutations may have their joint effect increased or diminished ($ W_{AB}$ larger or smaller), while in {\em (b)} the independent effect of the two mutations is deleterious and beneficial, respectively, and the combined expected effect on fitness is deleterious. In {\em (c)}, each mutation by itself is deleterious, but when they interact the result can be reciprocal sign epistasis (green arrow).  These sketches illustrate an additive model, where the sum of $W_A$ and $ W_B$ is equal to $ W_{AB}$ without epistasis. In our model using the geometric mean this corresponds to taking the logarithms of the fitness.} \label{epistasis_diagram}
\end{figure*}

Two mutations (A and B) occurring on a haplotype with wild-type fitness $W_0$ are said to be independent if the fitness effect of the joint mutation equals the product of the fitness effect of each of the mutations alone. If the fitness effect of the double mutant is $W_{AB}/W_0$ while the fitness effect of each of the single mutations is $W_A/W_0$ and $W_B/W_0$ respectively, then mutational independence implies (see illustration in Fig.~\ref{epistasis_diagram})
\be
\frac{W_A}{W_0}\frac{W_B}{W_0} = \frac{W_{AB}}{W_0}.
\ee
We quantify epistasis as the {\em deviation} from this equality, such that
\be
\varepsilon = \log \frac{W_0W_{AB}}{W_AW_B} \label{epsilon}
\ee
is zero when the combined effect of the two mutations is the same as the product of the individual effects on fitness. This definition is equivalent to the usual quantitative definition of epistasis in a two-locus two-allele model~(cf.~\cite{Bonhoeffer2004}, but see~\cite{Phillips2008} for a different definition) and transforms to the well-known additive definition of epistasis when the individual fitness effects are replaced by their logarithms (see, e.g.,~\cite{Mani2008}). Such a quantitative measure of epistasis was also used in assessing epistasis between mutations in experiments with {\it E. coli}~\cite{Elena1997} and digital organisms~\cite{Lenski1999}. For organisms on the line of descent (LOD, see Methods) of an evolutionary run, $A$ and $B$ refer to two substitutions that need not be adjacent either on the LOD or on the haplotype. We do see examples of non-consecutive mutations interacting, such as when a valley is crossed in more than one step (e.g., in Fig.~\ref{NK_case}, $K=4$), but here we restrict ourselves to studying the interaction between adjacent mutations on the LOD only, so that if $W_{AB}$ is the fitness of the haplotype that has both substitutions $A$ and $B$, then the type preceding this sequence on the LOD has fitness $W_A$. $W_B$ is found by reverting the first substitution ($A$), and measuring the fitness of the haplotype carrying only the second mutation ($B$). The relationship between the terms ``positive/negative" and ``synergistic/antagonistic" epistasis nomenclature is explained in Table 1.

\begin{table*}[!htbp]
\caption{Relationship between positive/negative and synergistic/antagonistic epistasis for different mutational pairs. Positive ($\varepsilon>0$) and negative ($\varepsilon<0$) epistasis imply synergistic/antagonistic if the two mutations are both beneficial or both deleterious, but when the mutations are of opposite effect the meaning of synergy or antagony is unclear (dashes). A  substitution can be characterized by how it interacts with the mutation that precedes it on the LOD using the {\em sign} of $\varepsilon$. Beneficial substitutions are designated B$^+$ or B$^-$, depending on whether they interacted epistatically with the preceding substitution to form positive or negative epistasis, respectively. D$^+$ and D$^-$ similarly indicate deleterious substitutions with positive and negative epistasis. Alternatively, writing BB$^+$ indicates that both substitutions increased fitness, and that the second substitution had a larger beneficial effect on the background of the first than it would have had on the background of the wild-type. DB$^-$ denotes a deleterious followed by a beneficial substitution that did not increase fitness as much as it would have if the deleterious substitution had not occurred.}
\vskip 1cm
\begin{tabular}{|cccc|cccc|@{}}
  \hline
 Designation & Epistasis & Sign & Effect &  Designation & Epistasis & Sign & Effect \\ \hline
 DD$^-$ & $\varepsilon<0$ & negative & synergistic & DB$^-$ & $\varepsilon<0$ & negative & --\\
  DD$^+$  & $\varepsilon>0$ & positive & antagonistic & DB$^+$ & $\varepsilon>0$ & positive & --\\\hline
  BB$^-$  & $\varepsilon<0$ & negative & antagonistic & BD$^-$ & $\varepsilon<0$ & negative & --\\
  BB$^+$ & $\varepsilon>0$ & positive & synergistic & BD$^+$ & $\varepsilon>0$ & positive & --\\\hline
\end{tabular}
\label{epidefs}
\end{table*}

\section*{Results}
\begin{figure}[h]
\includegraphics[width=0.5\textwidth]{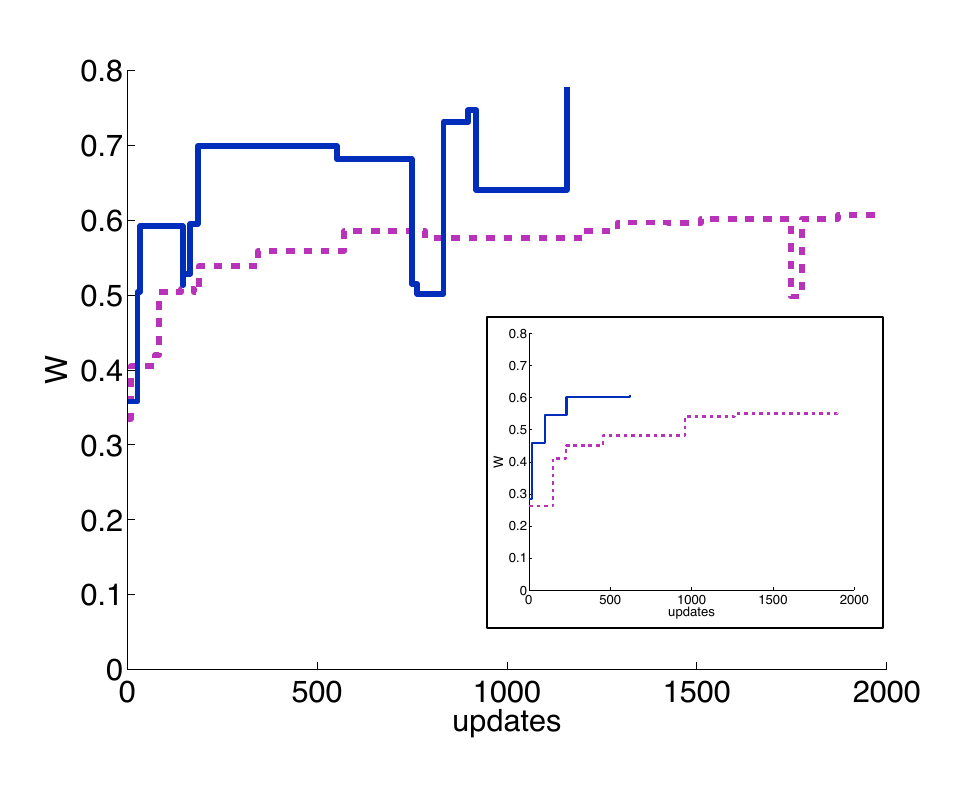}
\caption{Representative examples of adaptation in single lineages. Fitness on the line of descent for a simulation lasting 2,000 updates. The adaptive ascent is only shown until the lineage has attained the same fitness as it has after 2,000 updates, for $ N=20$, $ K=0$ (dashed) and $ K=4$ (solid),  at a high mutation rate $ \mu=10^{-2}$. The inset shows an example line of descent at $ \mu=10^{-4}$, with only beneficial mutations on the LOD.} \label{NK_case}
\end{figure}

We studied the impact of epistasis on adaptation by conducting evolutionary runs with different $K$ (which changes the landscape's ruggedness) and a fixed number of loci ($N=20$), for different mutation rates $\mu$ at a constant population size $\mathcal{N}$ of 5,000 individuals. In order to study only that part of evolutionary adaptation where a population climbs a local peak, each evolutionary run was initiated with a random selection of haplotypes of less than average fitness so that initially many beneficial mutations are available, akin to experiments with RNA viruses that are forced through bottlenecks~\cite{BurchChao1999,BurchChao2000}, or are subject to environmental change~\cite{Wichman1999}. The evolutionary dynamics of each run were similar in most cases: the population quickly adapts and situates itself near the top of a local peak, after which the population enters a period of stasis when exploration of the adjacent parts of the landscape does not turn up any more beneficial mutations (Fig.~\ref{NK_case}). This protocol is different (in terms of adaptation) from experiments in which only deleterious effects of mutations are studied, and advantageous mutations are found to be rare~\cite{Eyre-walker2007}. Thus, in this work we study the transient period of adaptation as opposed to mutation-selection balance. Initiating populations with a single genotype only does not change the evolutionary dynamics we observe (electronic supplementary material figure S1). 

\subsection*{Epistatic pairs on the LOD}
The mode of fixation of mutations, that is whether they go to fixation one by one or whether multiple mutations can interact in the same individual before fixation is determined by the mutation supply rate, i.e., the product of the mutation rate per genome per generation and the population size, $ \mu\mathcal{N}$. When this product is less than one, mutations usually go to fixation or are lost before the next mutation occurs~\cite{vanNimwegenetal1999,Campos2002,Gillespie2004,KimOrr2005}. This does not imply that mutations cannot interact, but instead that deleterious mutations are unlikely to be incorporated into the genome, because in that case all mutations must at least be neutral. When the mutation supply rate is significantly larger than one, mutations occur frequently enough that they can interact with each other before the first goes to fixation (the {\em concurrent mutations} regime~\cite{Desaietal2007,DesaiFisher2007,Fogleetal2008}). Given our population size of 5,000 and 20 loci, at the smallest rate we investigate ($\mu=10^{-4}$) the mutation supply rate is 10, but mutations still largely go to fixation separately with the result that the fraction of deleterious substitutions is less than one percent. The mutation supply rates we investigate range from 10 to 1,000, but are substantially smaller than the supply rate in the long-term evolution experiment with {\it E. coli} for example, because in that case the mutation rate per genome is $\mu\approx10^{-3}$~\cite{Barricketal2009} while the effective population size is $\mathcal{N}\approx 3\times 10^7$~\cite{Lenskietal1991}, for a $\mu\mathcal{N}\approx30,000$. 
While the per-locus mutation rate we use is higher than what we would expect in organisms that do not express a mutator phenotype~\cite{ZeyldeVisser2001}, we expect that the results will not change significantly if we could decrease the mutation rate while at the same time increasing the population size commensurately. In fact, it was shown (at least for neutral evolution~\cite{vanNimwegenetal1999}) that evolutionary dynamics is essentially unchanged if the two factors $\mu$ and ${\mathcal N}$ are  varied independently, as long as the product is the same. 

When the mutation supply rate is low, we do not expect that epistasis between mutations plays a significant role in the fixation of any individual mutation, simply because it is unlikely that any pair went to fixation in tandem. As a consequence, we expect that the number of interacting pairs on the line of descent of populations evolving at low mutation rate equals the rate at which they were produced. In other words, selection cannot amplify or reduce the number of interacting pairs. It is easy to compute how many pairs of mutations interact by chance in the NK model. If we ignore ``self-interactions" (a mutation can interact with itself when it is reversed by the next mutation on the LOD) the fitness of each locus is determined by $K$ others, but also plays a role in the fitness determination of $K$ other sites. As a consequence, $2K$ pairs out of the possible $N-1$ pairs (each locus can potentially interact with $N-1$ others in the absence of reversals) are interacting due to chance alone, that is, simply because they were within $K$ of each other. If we find more than $2K/(N-1)$ mutational pairs on the LOD that interact epistatically,  then we can conclude that these interactions contributed to why such pairs are on the LOD, in other words, that epistasis is selected for.

\begin{figure}[h]
\includegraphics[width=0.5\textwidth]{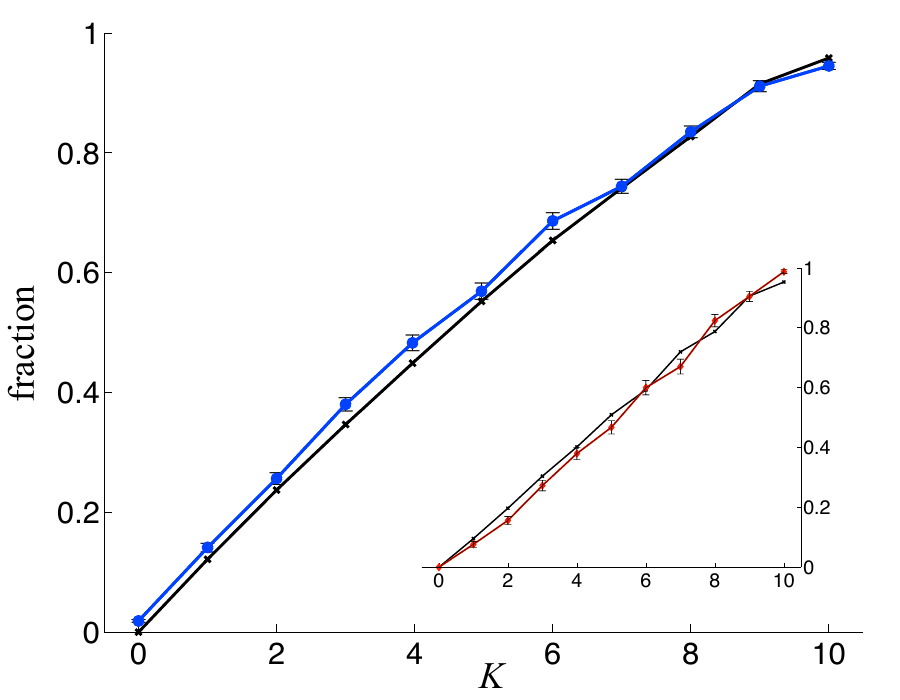}
\caption{Fraction of epistatic pairs on line of descent. The fraction of mutational pairs on the LOD that interact epistatically (circles) is larger than the numerical pre-selection prediction (crosses) for $\mu=10^{-2}$ ($p=0.013672$, Wilcoxon signed rank test). For smaller mutation rates ($\mu=10^{-4}$ shown in inset), there is no significant difference from the expectation ($p=0.23242$, Wilcoxon signed rank test). Lines are drawn to guide the eye.} \label{pairs_lod_theory}
\end{figure}

Given $N$, $K$, and $\mu$, we first numerically compute the fraction of all mutational pairs that will interact {\em before} the mutations are screened by selection, by randomly mutating a haplotype and testing if any mutations are a distance of $K$ loci or less away from each other. 
Even though the null expectation is $K/(N-1)$, we perform this numerical estimate because the fraction also depends on the mutation rate: The higher the mutation rate, the greater the chance that an individual haplotype will be hit by more than one mutation, which elevates the fraction of available interacting mutations above $K/(N-1)$.  We found that the fraction of epistatic pairs on the LOD differs significantly from the fraction available (the pre-selection prediction) when the mutation rate is high ($\mu=10^{-2}$, Fig.~\ref{pairs_lod_theory}). Because deleterious mutations enable organisms to cross valleys between peaks, the LOD is enriched by epistatic pairs that include deleterious mutations. For smaller mutation rates this is not the case, as valleys cannot be crossed (see inset in Fig.~\ref{pairs_lod_theory} for $\mu=10^{-4}$). 

\subsection*{Mechanism of interaction between mutations}
\begin{figure}[htbp] 
   \centering
   \includegraphics[width=0.5\textwidth]{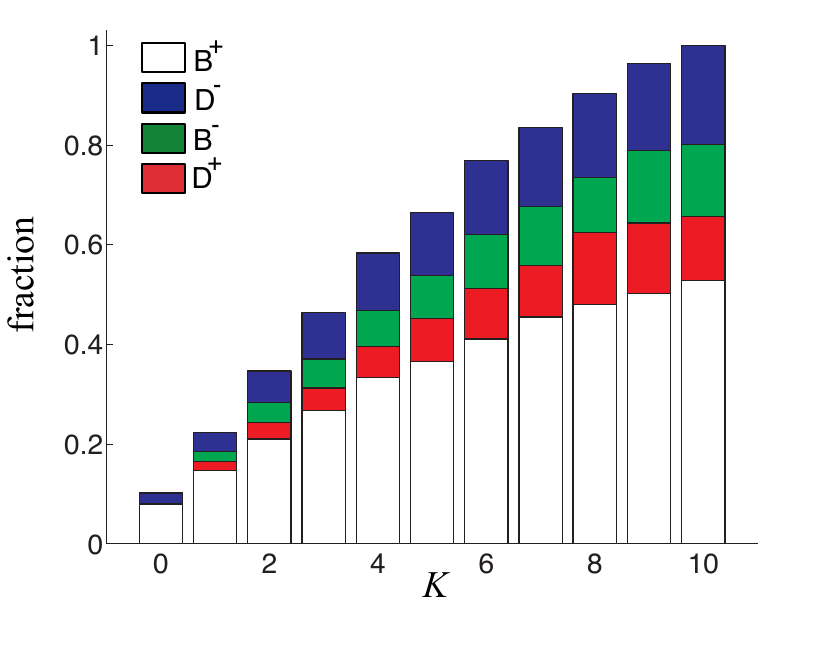}
\caption{Fraction of types of the second substitution among all epistatic pairs.
Height of bar shows the fraction of all epistatic mutation of a particular type on the LOD.  At the highest mutation rate tested $(\mu=10^{-2}$), a considerable fraction of epistatic substitutions are D$^+$ and D$^-$, while those fractions are less at lower mutation rates (see legend for colour-type assignment). 
   \label{fig:frac-types}}
\end{figure}

Beneficial as well as deleterious mutations can interact positively or negatively, as summarized in Table 1.
The majority of consecutive pairs of mutations on the LOD are pairs of beneficial mutations (BB pairs, see supplementary electronic material figure S3A), followed by BD, DB, and DD pairs. The relative fraction of these pairs depends on the mutation rate, but is roughly independent of $K$. How (and how often) these mutations interact, however, does depend on $K$. Let us first look at the second mutation of an interacting pair of substitutions, to which we can give the labels B$^+$, B$^-$, D$^+$, and D$^-$, depending on whether they were beneficial or deleterious on the background of the preceding mutation, and on whether they interacted positively or negatively with it. 
If they interact, substitutions show positive epistasis with the mutation preceding them on the LOD, as we would expect for substitutions accumulated on a fitness ascent, as we can see in Fig.~\ref{fig:frac-types}.  All four types of epistatic mutations increase in frequency at the expense of mutations that do not interact epistatically. Overall, we observe that as $K$ increases, the population uses deleterious mutations that interact epistatically  to adapt more efficiently, as valleys are crossed to ascend higher fitness peaks. This effect is severely diminished when the mutation supply rate is low ($\mu \mathcal{N}<10$), in which case mutations typically go to fixation before a second mutation occurs (supplementary electronic material figure S2).
Crossing fitness barriers is enabled mostly by pairs of the type DB$^+$, that is, a deleterious mutation followed by a mutation whose benefit is enhanced by the presence of the preceding deleterious mutation (see supplementary electronic material figure S3B). This synergy between deleterious and beneficial mutations can go as far as {\it sign epistasis}, that is, a mutation that is only beneficial in the presence of the preceding deleterious, but deleterious in the absence of it.
At $K=0$ most mutational pairs consist of two beneficial mutations that do not interact epistatically, except when the second mutation occurs at the same locus as the first, thereby reversing the first mutation (see Fig.~\ref{fig:frac-types}).  
Reversals mostly consist of deleterious-beneficial pairs exhibiting positive epistasis DB$^+$, with a small minority of beneficial-deleterious pairs exhibiting negative epistasis (BD$^-$). Once $K=5$, most of the DB pairs show positive epistasis, showing that interacting mutations ease the traversal of fitness barriers (supplementary electronic material figure S3).

\begin{figure*}[!htbp]
\center  \includegraphics[width=\textwidth]{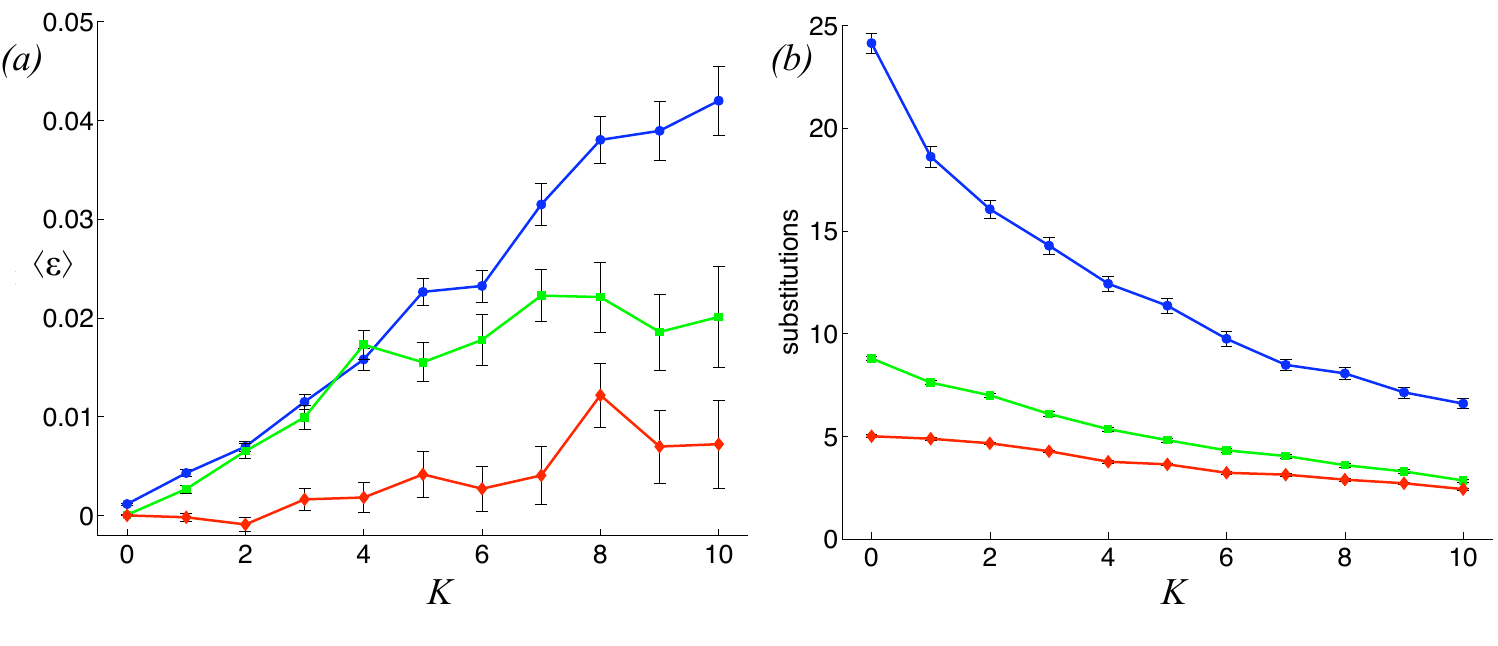}
\caption{Mean $\varepsilon$ and substitutions on the line of descent. {\em (a)} $\la\varepsilon\ra$ on the LOD as defined by Eq.~(\ref{meanepsilon}).  Each datum is the average of 200 LODs and error bars are standard error. Mutation rates are $ \mu=10^{-2}$ (blue circles), $ \mu=10^{-3}$ (green squares), and $\mu=10^{-4}$ (red diamonds). Population size is 5,000, $ N=20$, and the replacement rate is 10\%. Lines are drawn to guide the eye.  {\em (b)}: Total number of substitutions as a function of $K$,  mutation rates and colours as in {\em (a)}. }\label{epi_mu}
\end{figure*}
\subsection*{Correlation between epistasis and beneficial effect}
We define the \emph{mean size of epistasis on the LOD} as the mean $\varepsilon$ between all consecutive pairs:
\be
\la\varepsilon\ra \equiv \frac1n\sum_{i=1}^n \varepsilon_i\;, \label{meanepsilon}
\ee
where the sum runs over all substitutions on the LOD, $\varepsilon_i$ is the size of epistasis of the $i$th pair [between mutation $i+1$ and $i$ on the LOD, given by Eq.~(\ref{epsilon})], and $n$ is the number of pairs (one less than the number of substitutions).  This measure has an expectation value of zero if negatively and positively interacting pairs occur with equal likelihood, and with equal and opposite strength, on the LOD. We are studying the mean of $\varepsilon$ in order to compare this measure across evolutionary runs that differ in the average number of mutations on the LOD. We find that $\la\varepsilon\ra$ increases with $K$  for all three mutation rates (Fig.~\ref{epi_mu}a). Higher mutation rates result in larger $\la\varepsilon\ra$  on the LOD, because the higher rate decreases the waiting time for new mutations, making it easier for a lineage to cross a valley in the fitness landscape via a deleterious mutation. If a mutation is deleterious, the lineage that carries this mutation needs another mutation that at least compensates for the fitness loss before the lineage goes extinct.

\begin{figure}[h]
 \includegraphics[width=0.5\textwidth]{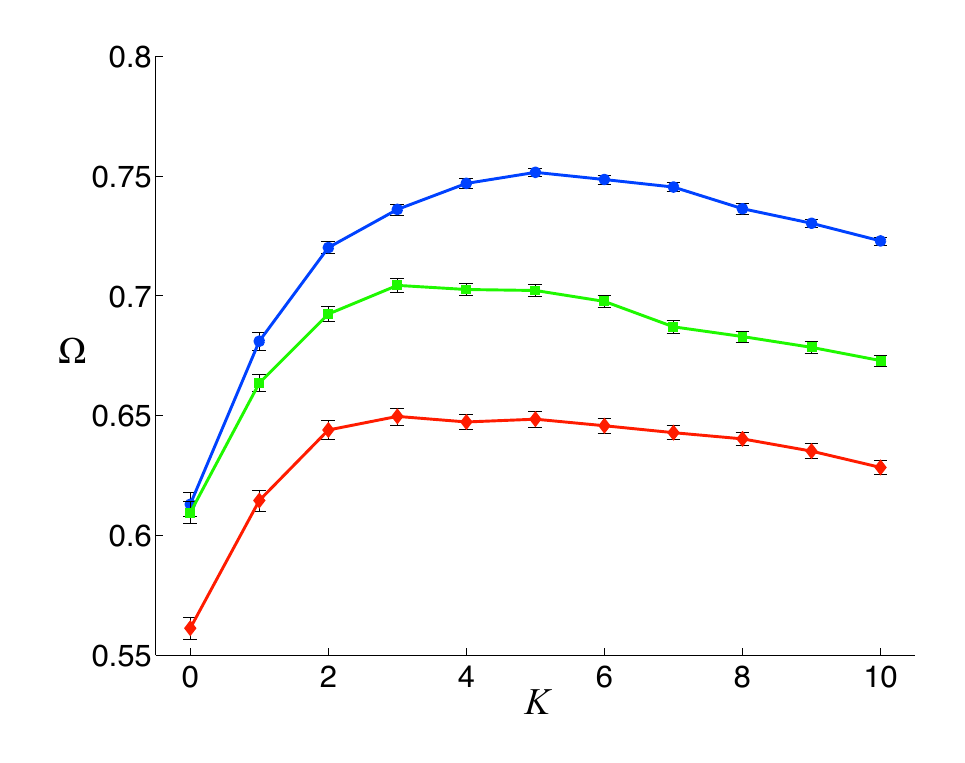}
\caption{Attained fitness $ \Omega$,as a function of $ K$ for three different mutation rates (colours and parameters as in Fig.~\ref{epi_mu}a) on LOD. $ K_{\rm opt}$, the point at which $ \Omega$ is maximal, is larger for higher mutation rates.} \label{omega_mu}
\end{figure}

While $\la\varepsilon\ra$ increases with $K$, the number of substitutions during adaptation decreases (Fig.~\ref{epi_mu}b), and the fraction of deleterious substitutions is mostly unchanged between low and intermediate $K$ (supplementary electronic material figure S4). The origin of the decrease in the number of substitutions is clear: for $K=0$, mutations that increase fitness are not difficult to find because the landscape is smooth. More rugged landscapes risk confining the population to local peaks, and even though valleys can be crossed towards higher fitness peaks that are close, ultimately the ruggedness puts a stop to further adaptation~\cite{Weissman2009}. Even though the number of substitutions decreases with $K$, higher fitness levels are achieved at intermediate $K$ compared to lower $K$. Indeed, the attained fitness, $\Omega$ (the fitness of the best genotype at the end of a simulation run), increases with $K$ up to intermediate values (Fig.~\ref{omega_mu}a), and the time to reach the attained fitness is shorter the higher $K$ is (supplementary electronic material figure S5).  This also explains why the observed attained fitness for $K=0$ and $\mu=10^{-4}$ is not maximal in Fig.~\ref{omega_mu}a as we would expect for a smooth landscape. Increasing the simulation time to 100,000 updates does give the population enough time to reach the peak.
For $K\leq5$ the attained fitness  is an increasing function of both $\langle\varepsilon\rangle$ and the mean selection coefficient (supplementary electronic material figure S6), that is, higher $\langle\varepsilon\rangle$ goes hand in hand with higher achieved fitness.

\begin{figure}[!htbp]
 \includegraphics[width=0.5\textwidth]{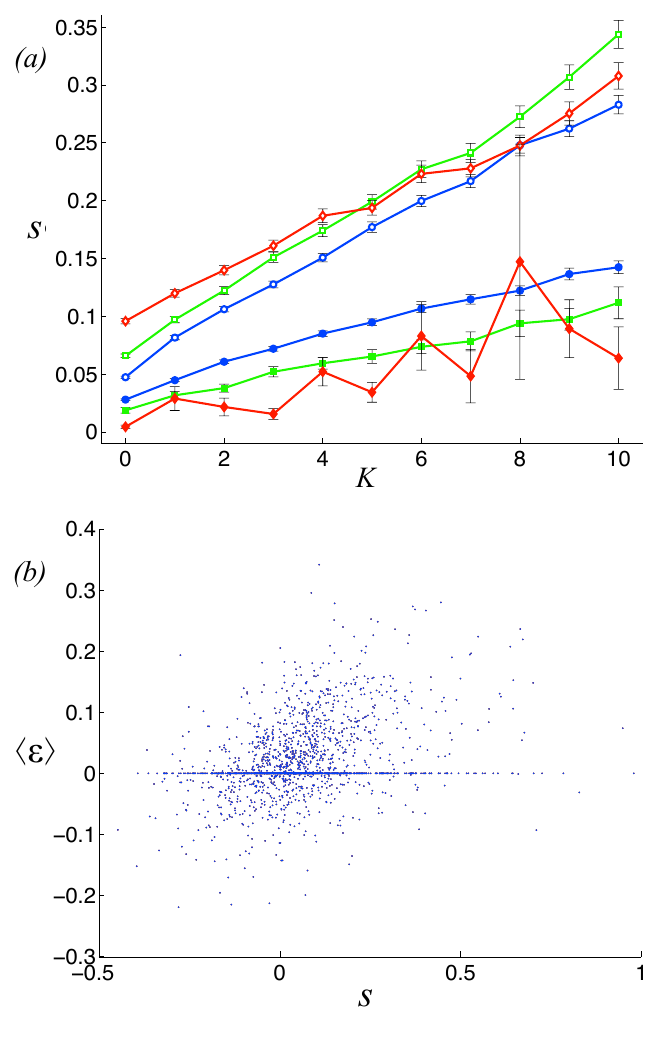}
\caption{Strength of selection coefficients and epistasis. {\em (a)}: The effect on fitness of beneficial, $ s_b$ (open symbols), and deleterious substitutions, $ s_d$ (solid symbols), both increase approximately linearly as a function of $ K$ (colours and parameters as in Fig.~\ref{epi_mu}). {\em (b)}: Correlation between size of epistasis $\la\varepsilon\ra$ and effect of substitutions $s$, shown here for $K=5$  and $ \mu=10^{-2}$. Reversal substitutions are excluded because they do not contribute to adaptation. Including them would only strengthen the overall correlation. Pearson correlation coefficient $r=0.3549$.} \label{s_mu}
\end{figure}

That higher fitness can be achieved with fewer substitutions seems counterintuitive (but see~\cite{MacLean2009}), yet is an effect achieved both by epistasis and pleiotropy. Pleiotropy can result in a single mutation increasing $K+1$ fitness components {\em at the same time}, leading to the same fitness increase with fewer mutations. With luck, one mutation will increase fitness in all or most of the $K+1$ components that it affects, amplifying the effect of the mutation. Pleiotropy is therefore directly responsible for the increase in potential selection coefficients as a function of $K$. Even though the chance that a mutation will have a positive effect on all $K+1$ interacting loci becomes smaller as $K$ increases, the relationship between fitness increase per substitution is an approximately linear function of $K$ (Fig.~\ref{s_mu}), indicating that each mutation on the LOD carries a ``bigger punch" as the number of interacting loci, $K$, increases. Both peak frequency and amplitude correlate with $K$, and together these two cause the increase in average selection coefficients for mutations by increasing  the {\em slope} leading up to the peaks. Just such an interaction between traits to achieve higher fitness has also recently been observed in quantitative trait loci affecting skeletal characters in mice~\cite{Wagneretal2008}. 

Besides changing the degree of pleiotropy, $K$ also directly modulates epistasis. More epistasis causes the frequency of peaks and valleys to increase, which, in addition to pleiotropy, causes increased selection coefficients. The correlation between the benefit a mutation provides and the amount of epistasis $\varepsilon$ between this and other mutations, as evidenced by Figs.~\ref{s_mu}a and b, mirrors the observation of a correlation between directional epistasis and the {\em deleterious} effects of mutations seen in other computational studies of evolution~\cite{WilkeAdami2001,WilkeLenskiAdami2003,Azevedoetal2006}, as well as in protein evolution \emph{in vitro} \cite{Bershteinetal2006},  bacterial evolution~\cite{Beerenwinkeletal2007}, and even viroids~\cite{Sanjuanetal2006}. Because beneficial mutations are rare in most of these studies, a correlation between positive effects and epistasis has not be shown before. Varying the mutation rate does not qualitatively change these results. 

As $K$ increases, the mean height of peaks decreases beyond $K=1$ (supplementary electronic material figure S7) because many more shallow peaks appear than high ones. Yet, the global peak height continues to increase beyond $K=5$ (but note that peaks can never exceed $W=1$, no matter what the $K$). Thus, part of the observed effect comes from the fact that landscapes with higher $K$ contain higher peaks~\cite{Skellettetal2005}. To test whether the observed increase in mean beneficial effect with $K$ of single substitutions solely stems from the increase in peak height, we ran control simulations in which the fitness landscape is normalized such that the range in fitness is the same across all $K$. In this case the global peak is fixed at the same height in all fitness landscapes, while the frequency of peaks remains unaffected. In this instance of the NK model, the attained fitness is never larger than what can be attained at $K=0$, and decreases as $K$ increases due to the increased ruggedness of the landscape (see supplementary electronic material figure S8A). Yet, the selection coefficients are still an increasing function of $K$ even in the normalized landscape, but the slope is shallower than for the non-normalized model  (supplementary electronic material figure S8B). Thus, beneficial mutations still cooperate synergistically for a ``bigger punch" per mutation, even if the peak height is normalized.

\section*{Discussion}
We studied how interacting mutations impact the evolutionary dynamics for populations evolving in an artificial fitness landscape in which the ruggedness is determined by a single parameter $K$, in the ``strong mutation" limit. 
We found that increasing the ruggedness of the landscape by raising $K$ has several consequences. First, the mean epistatic effect per substitution $\la\varepsilon\ra$ monotonically increases with $K$ (Fig.~\ref{epi_mu}). Second, the number of substitutions on the LOD decreases, while the fraction of those substitutions that are deleterious or beneficial remains largely unchanged (Fig.~S4). We might intuit that fewer beneficial substitutions will impair adaptation, but here we instead observe a third effect, namely that those higher peaks that appear as $K$ is increased can be located faster and in fewer steps (Fig.~\ref{omega_mu}), because the mean selection coefficient per mutation (Fig.~\ref{s_mu}) increases with $K$ in an approximately linear fashion. This effect is robust even if we correct for the increasing height of fitness peaks with increasing $K$ in this landscape, which reflects the proliferation of adaptive opportunities that come with synergistic interactions.

Ruggedness is normally viewed as an impediment to adaptation, because the presence of valleys means that the organism has to suffer a decrease in fitness before it can gain a fitness advantage~\cite{Clune2008}. However, in the NK model, increased ruggedness not only translates into more peaks to ascend and more valleys to cross, but also increases both the fitness {\em difference} between the peaks and valleys (amplitude) and the height of the global peak.  The attained fitness is maximal at $K=3\ {\rm to}\ 5$, from which we infer that an intermediate amount of epistasis and pleiotropy is most conducive to adaptation (Fig.~\ref{omega_mu}a). The population is able to take advantage of the presence of higher peaks that exist for higher $K$, particularly for the highest mutation rate. The observed {\em decrease} in attained fitness at high $K$ is caused by longer waiting times to new mutations (as was shown in~\cite{Clune2008}) which is a consequence of the increasingly rugged structure of the NK landscape for high $K$. As we increase $K$, the increased average effect of single mutations (either beneficial or deleterious) is counterbalanced by the increasing ruggedness of the landscape, which makes it more likely that the population becomes stuck on a suboptimal fitness peak instead of locating the global peak. This lowers the average attained fitness of the population compared to lower $K$. 

As the number of peaks increases with $K$ (thus shortening the mutational distance between peaks), the fitness decrease that an organism must endure while traversing the valley in-between the peaks becomes larger. As a consequence, successful lineages must have an increased benefit per substitution for higher $K$. Indeed, with increasing $K$ the distribution of single-mutation fitness effects becomes broader (supplementary electronic material figure S9), allowing some mutations to increase the fitness of the haplotype by as much as a factor of $K+1$ compared to the $K=0$ case. This is an effect of pleiotropy, which is inseparable from epistasis in this implementation of the NK model, and has a direct counterpart in empirical fitness landscapes as well~\cite{Wagneretal2008}. Further investigation into the different roles and impacts of epistasis and pleiotropy is important for the understanding of the dynamics of the NK model, as well as for the relative roles of epistasis and pleiotropy in adaptive evolution.

As discussed, the results presented here are a consequence of the increased frequency and amplitude of the peaks as well as the increase in global peak height (the height of the highest peak) of the landscape as $K$ is increased. It could be argued that this increase in the size of the highest peaks (even as the mean peak height decreases) is an artifact of the NK model that has no counterpart in how biological fitness landscapes change when the number of interactions between genes changes.  Instead, we believe that the increase in adaptive potential is germane, because interacting loci can work synergistically to produce higher fitness compared to a set of non-interacting loci. In a sense, increasing $K$ creates a more modular landscape of epistatically interacting genes. Indeed, searching for epistatically interacting genes is one method to search for modules in metabolic genes~\cite{segreetal2005}, and a clustering method has been used recently to find modules from epistatically interacting pairs of genes in yeast~\cite{Costanzoetal2010}. Those authors found a dependence of the fraction of pairs that are epistatic on the size of the deleterious effect of a mutation that mirrors the dependence we observe here (supplementary electronic material figure S10), and thus strongly epistatic pairs of mutations provide the largest fitness benefit also in yeast. 

For the NK model, we can understand why the global peak fitness increases as a direct consequence of the modularity of the fitness components: each fitness component $w_i$ is controlled by $K+1$ loci, giving $2^{K+1}$ possible values. It is more likely to find higher fitness values in those larger samples. So, just as in the biological pathways with modular structure in yeast, the more loci that contribute to a fitness component, the better this component can be fine-tuned to optimize its contribution to fitness. Given these considerations, we contend that the NK fitness landscape, obtained from interacting loci that synergistically contribute to the function of traits, is a reasonable and appropriate model for describing interacting gene networks in biological organisms. 

\section*{Methods}
\subsection*{Simulations}
We simulated the evolutionary process by randomly removing 10\% of the population every update, and replacing them with copies of a subset of the remainders, selected with probabilities proportional to individual fitness. This is akin to the Wright-Fisher model for haploid asexuals~\cite{DonnellyWeber1985}, but with overlapping generations. In evolution experiments implemented in flow reactors (for example, continuous culture experiments, see~\cite{Lindemannetal2002}), the replacement rate is akin to the flow rate of the reactor.  Varying the replacement rate does not change the conclusions we reach in this study. We define the period of adaptation as beginning at update zero, and ending when the lineage first reaches the same fitness that it acquired at the end of the simulation. In this manner, we exclude from the analysis reversal mutations (i.e., mutations undoing previous mutations at the same locus) that occasionally occur {\em after} a fitness peak has been ascended. If we included those reversals, both the number of deleterious substitutions and the amount of epistasis measured would be affected, even though they do not contribute to adaptation. 

In order to study the part of the evolutionary trajectory that corresponds to climbing the nearest fitness peak, we choose as the ancestral population a sample of individuals with fitness in the lowest 50\% of a randomly generated population where the haplotypes of the individuals are uncorrelated. As a consequence, many beneficial mutations are possible, so individual lineages may climb different peaks (except for $K=0$, in which case there is only one peak), and the lineage that happens to climb the fastest will be most likely to outcompete the other organisms in the population. This protocol is similar in spirit to that used in references~\cite{BurchChao1999,BurchChao2000}, where a population of $\Phi$6 viruses was put through bottlenecks in order to study the dynamics of re-adaptation. For each mutation rate and for each $K$, we collected 200 independent evolutionary runs and extracted one line of descent (LOD, see below) from each. In our results, we report the average values across these 200 samples, and provide standard errors. 

The probability of each locus changing its binary value is set by a per-site mutation rate $\mu$. While the average rate of mutation is fixed, the process itself is stochastic so that the distribution of the number of mutations per organism is Poisson-random with the given mean. We varied $K$ from 0 (no interaction between neighboring loci) to 10, where each locus interacts with ten of its neighboring loci.  Because the haplotypes are circular, for $K=10$ all mutational pairs interact (100\% of mutational pairs are epistatic).

\subsection*{Line of Descent}
We study the sequence of mutations that accumulates as populations adapt from an initial state of low fitness to the maximum fitness they can attain given their environment, by studying a single individual lineage from its inception to the end of the simulation run (typically 2,000 updates of the population). We do this by picking the most fit organism after a set number of simulation updates, and then track this individual's ancestry all the way back to the beginning of the simulation. We define this sequence of mutations as the line of descent (LOD), and discard all other data from that simulation~\cite{Lenskietal2003,Cowperthwaite2006}. For asexual populations in a single niche (no frequency-dependent selection), the LOD accurately represents the population as each substitution that appears on the LOD must be shared by the entire population by definition, from the most common recent ancestor on all the way to the origin.

\section*{Acknowledgments}
This work was supported in part by a grant from the Cambridge Templeton Consortium,  by the National Science Foundation's Frontiers in Integrative Biological Research grant FIBR-0527023, and by NSF's BEACON Center for the Study of Evolution in Action, under cooperative agreement No. DBI-0939454. The funders had no role in study design, data collection and analysis, decision to publish, or preparation of the manuscript.


\begin{thebibliography}{10}
\expandafter\ifx\csname urlstyle\endcsname\relax
  \providecommand{\doi}[1]{doi:\discretionary{}{}{}#1}\else
  \providecommand{\doi}{doi:\discretionary{}{}{}\begingroup
  \urlstyle{rm}\Url}\fi

\bibitem{Fisher1930}
Fisher, R.  1930 \emph{The Genetical Theory of Natural Selection}.
\newblock Oxford, UK: Oxford University Press.

\bibitem{BurchChao1999}
Burch, C.~L. \& Chao, L.  1999 Evolution by small steps and rugged landscapes
  in the {RNA} virus {$\Phi$6}.
\newblock \emph{Genetics} \textbf{151}, 921--927.

\bibitem{Orr2005}
Orr, H.~A.  2005 The genetic theory of adaptation: A brief history.
\newblock \emph{Nature Reviews Genetics} \textbf{6}, 119--127.
\newblock \doi{10.1038/nrg1523}.

\bibitem{Gillespie1984}
Gillespie, J.~H.  1984 Molecular evolution over the mutational landscape.
\newblock \emph{Evolution} \textbf{38}, 1116--1129.

\bibitem{Gillespie1991}
Gillespie, J.  1991 \emph{The Causes of Molecular Evolution}.
\newblock New York, NY: Oxforn University Press.

\bibitem{Orr2002}
Orr, H.  2002 The population genetics of adaptation: The adaptation of {DNA}
  sequences.
\newblock \emph{Evolution} \textbf{56}, 1317--1330.
\newblock \doi{10.1111/j.0014-3820.2002.tb01446.x}.

\bibitem{KimOrr2005}
Kim, Y. \& Orr, H.~A.  2005 Adaptation in sexuals vs. asexuals: clonal
  interference and the fisher-muller model.
\newblock \emph{Genetics} \textbf{171}, 1377--86.
\newblock \doi{10.1534/genetics.105.045252}.

\bibitem{Kryazhimskiyetal2009}
Kryazhimskiy, S., Tkacik, G. \& Plotkin, J.~B.  2009 The dynamics of adaptation
  on correlated fitness landscapes.
\newblock \emph{Proc Natl Acad Sci U S A} \textbf{106}, 18,638--43.
\newblock \doi{10.1073/pnas.0905497106}.

\bibitem{Kimura1985}
Kimura, M.  1985 The role of compensatory neutral mutations in molecular
  evolution.
\newblock \emph{Journal of Genetics} \textbf{64}, 7--19.
\newblock \doi{10.1007/BF02923549}.

\bibitem{Lenskietal2003}
Lenski, R.~E., Ofria, C., Pennock, R.~T. \& Adami, C.  2003 The evolutionary
  origin of complex features.
\newblock \emph{Nature} \textbf{423}, 139--144.
\newblock \doi{10.1038/nature01568}.

\bibitem{Bridgham2006}
Bridgham, J.~T., Carroll, S.~M. \& Thornton, J.~W.  2006 Evolution of
  hormone-receptor complexity by molecular exploitation.
\newblock \emph{Science} \textbf{312}, 97--101.
\newblock \doi{10.1126/science.1123348}.

\bibitem{Poelwijk2006}
Poelwijk, F.~J., Kiviet, D.~J. \& Tans, S.~J.  2006 Evolutionary potential of a
  duplicated repressor-operator pair: Simulating pathways using mutation data.
\newblock \emph{PLoS Computational Biology} \textbf{2}, 467--475.
\newblock \doi{10.1371/journal.pcbi.0020058}.

\bibitem{Cowperthwaite2006}
Cowperthwaite, M.~C., Bull, J.~J. \& Meyers, L.~A.  2006 From bad to good:
  Fitness reversals and the ascent of deleterious mutations.
\newblock \emph{PLoS Computational Biology} \textbf{2}, 1292--1300.
\newblock \doi{10.1371/journal.pcbi.0020141}.

\bibitem{Clune2008}
Clune, J., Misevic, D., Ofria, C., Lenski, R.~E., Elena, S.~F. \& Sanju\'an, R.
   2008 Natural selection fails to optimize mutation rates for long-term
  adaptation on rugged fitness landscapes.
\newblock \emph{PLoS Computational Biology} \textbf{4}, e1000,187.
\newblock \doi{10.1371/journal.pcbi.1000187}.

\bibitem{ZuckerkandlPauling1965}
Zuckerkandl, E. \& Pauling, L.  1965 Evolutionary divergence and convergence in
  proteins.
\newblock In V.~Bryson \& H.~J. Vogel, eds., \emph{Evolving Genes and Proteins}
   pp. 97--166. Academic Press.

\bibitem{BloomArnold2009}
Bloom, J.~D. \& Arnold, F.~H.  2009 {In the light of directed evolution:
  Pathways of adaptive protein evolution}.
\newblock \emph{Proceedings of the National Academy of Sciences of the United
  States of America} \textbf{106}, 9995--10,000.
\newblock \doi{10.1073/pnas.0901522106}.

\bibitem{Phillips2008}
Phillips, P.~C.  2008 Epistasis - the essential role of gene interactions in
  the structure and evolution of genetic systems.
\newblock \emph{Nature Reviews Genetics} \textbf{9}, 855--867.
\newblock \doi{10.1038/nrg2452}.

\bibitem{WeinreichWatsonChao2005}
Weinreich, D.~M., Watson, R.~A. \& Chao, L.  2005 Sign epistasis and genetic
  constraint on evolutionary trajectories.
\newblock \emph{Evolution} \textbf{59}, 1165--1174.
\newblock \doi{10.1111/j.0014-3820.2005.tb01768.x}.

\bibitem{Poelwijk2007}
Poelwijk, F.~J., Kiviet, D.~J., Weinreich, D.~M. \& Tans, S.~J.  2007 Empirical
  fitness landscapes reveal accessible evolutionary paths.
\newblock \emph{Nature} \textbf{445}, 383--386.
\newblock \doi{10.1038/nature05451}.

\bibitem{Reetzetal2005}
Reetz, M.~T., Bocola, M., Carballeira, J.~D., Zha, D.~X. \& Vogel, A.  2005
  Expanding the range of substrate acceptance of enzymes: Combinatorial
  active-site saturation test.
\newblock \emph{Angewandte Chemie-International Edition} \textbf{44},
  4192--4196.
\newblock \doi{10.1002/anie.200500767}.

\bibitem{WeinreichChao2005}
Weinreich, D.~M. \& Chao, L.  2005 Rapid evolutionary escape by large
  populations from local fitness peaks is likely in nature.
\newblock \emph{Evolution} \textbf{59}, 1175--1182.
\newblock \doi{10.1111/j.0014-3820.2005.tb01769.x}.

\bibitem{Weinreich2006}
Weinreich, D.~M., Delaney, N.~F., DePristo, M.~A. \& Hartl, D.~L.  2006
  Darwinian evolution can follow only very few mutational paths to fitter
  proteins.
\newblock \emph{Science} \textbf{213}, 111--114.
\newblock \doi{10.1126/science.1123539}.

\bibitem{Lockless1999}
Lockless, S.~W. \& Ranganathan, R.  1999 Evolutionarily conserved pathways of
  energetic connectivity in protein families.
\newblock \emph{Science} \textbf{286}, 295--299.
\newblock \doi{10.1126/science.286.5438.295}.

\bibitem{Whitlock1995}
Whitlock, M.~C., Phillips, P.~C., Moore, F. B.-G. \& Tonsor, S.~J.  1995
  Multiple fitness peaks and epistasis.
\newblock \emph{Annual Review of Ecology and Systematics} \textbf{26}, 601--29.

\bibitem{Coyneetal2000}
Coyne, J.~A., Barton, N.~H. \& Turelli, M.  2000 Is {Wright's} shifting balance
  process important in evolution?
\newblock \emph{Evolution} \textbf{54}, 306--317.
\newblock \doi{10.1111/j.0014-3820.2000.tb00033.x}.

\bibitem{Philippsetal2000}
Phillips, P., Otto, S. \& Whitlock, M.  2000 Beyond the average, the
  evolutionary importance of gene interactions and variability of epistatic
  effects.
\newblock In J.~Wolf, E.~Brodie~III \& M.~Wade, eds., \emph{Epistasis and the
  Evolutionary Process}  pp. 20--38. Oxford University Press.
\newblock \doi{Add data for field: Doi}.

\bibitem{KelleyIdeker2005}
Kelley, R. \& Ideker, T.  2005 Systematic interpretation of genetic
  interactions using protein networks.
\newblock \emph{Nature Biotechnology} \textbf{23}, 561--566.
\newblock \doi{10.1038/nbt1096}.

\bibitem{UlitskyShamir2007}
Ulitsky, I. \& Shamir, R.  2007 Pathway redundancy and protein essentiality
  revealed in the \it {Saccharomyces cerevisiae} \rm interaction networks.
\newblock \emph{Molecular Systems Biology} \textbf{3}, 104.
\newblock \doi{10.1038/msb4100144}.

\bibitem{Roguevetal2008}
Roguev, A., Bandyopadhyay, S., Zofall, M., Zhang, K., Fischer, T., Collins,
  S.~R., Qu, H., Shales, M., Park, H.-O., Hayles, J. \emph{et~al.}  2008
  Conservation and rewiring of functional modules revealed by an epistasis map
  in fission yeast.
\newblock \emph{Science} \textbf{322}, 405--410.
\newblock \doi{10.1126/science.1162609}.

\bibitem{Costanzoetal2010}
Costanzo, M., Baryshnikova, A., Bellay, J., Kim, Y., Spear, E.~D., Sevier,
  C.~S., Ding, H., Koh, J. L.~Y., Toufighi, K., Mostafavi, S. \emph{et~al.}
  2010 The genetic landscape of a cell.
\newblock \emph{Science} \textbf{327}, 425--31.
\newblock \doi{10.1126/science.1180823}.

\bibitem{KauffmanLevin1987}
Kauffman, S. \& Levin, S.  1987 Towards a general theory of adaptive walks on
  rugged landscapes.
\newblock \emph{Journal of Theoretical Biology} \textbf{128}, 11--45.
\newblock \doi{10.1016/S0022-5193(87)80029-2}.

\bibitem{KauffmanWeinberger1989}
Kauffman, S.~A. \& Weinberger, E.~D.  1989 The {NK} model of rugged fitness
  landscapes and its application to maturation of the immune response.
\newblock \emph{Journal of Theoretical Biology} \textbf{141}, 211--245.
\newblock \doi{10.1016/S0022-5193(89)80019-0}.

\bibitem{Kauffman1993}
Kauffman, S.~A.  1993 \emph{The Origins of Order: Self-Organization and
  Selection in Evolution}.
\newblock Oxford University Press US.

\bibitem{Altenberg1997}
Altenberg, L.  1997 {NK} fitness landscapes.
\newblock In T.~Back, D.~Fogel \& Z.~Michalewicz, eds., \emph{The Handbook of
  Evolutionary Computation}  pp. B2.7:5--10. IOP Publishing.

\bibitem{MackenPerelson1989}
Macken, C.~A. \& Perelson, A.~S.  1989 Protein evolution on rugged landscapes.
\newblock \emph{Proceedings of the National Academy of Sciences of the United
  States of America} \textbf{86}, 6191--6195.

\bibitem{PerelsonMacken1995}
Perelson, A.~S. \& Macken, C.~A.  1995 Protein evolution on partially
  correlated landscapes.
\newblock \emph{Proceedings of the National Academy of Sciences of the United
  States of America} \textbf{92}, 9657--9661.

\bibitem{Solowetal1999}
Solow, D., Burnetas, A., Roeder, T. \& Greenspan, N.~S.  1999 Evolutionary
  consequences of selected locus-specific variations in epistasis and fitness
  contribution of {Kauffman's NK model}.
\newblock \emph{Journal of theoretical Biology} \textbf{196}, 181--196.
\newblock \doi{10.1006/jtbi.1998.0832}.

\bibitem{Campos2002}
Campos, P.~R., Adami, C. \& Wilke, C.~O.  2002 Optimal adaptive performance and
  delocalization in {NK} fitness landscapes.
\newblock \emph{Physica A} \textbf{304}, 495--506.
\newblock \doi{10.1016/S0378-4371(01)00572-6}.

\bibitem{WelchWaxman2005}
Welch, J.~J. \& Waxman, D.  2005 The {NK} model and population genetics.
\newblock \emph{Journal of Theoretical Biology} \textbf{234}, 329--340.
\newblock \doi{10.1016/j.jtbi.2004.11.027}.

\bibitem{Orr2006}
Orr, H.~A.  2006 The population genetics of adaptation on correlated fitness
  landscapes: The block model.
\newblock \emph{Evolution} \textbf{60}, 1113--1124.
\newblock \doi{10.1111/j.0014-3820.2006.tb01191.x}.

\bibitem{Wagner1996}
Wagner, A.  1996 Does evolutionary plasticity evolve?
\newblock \emph{Evolution} \textbf{50}, 1008--1023.
\newblock \doi{10.2307/2410642}.

\bibitem{Cilibertietal2007}
Ciliberti, S., Martin, O.~C. \& Wagner, A.  2007 Robustness can evolve
  gradually in complex regulatory gene networks with varying topology.
\newblock \emph{Plos Computational Biology} \textbf{3}, 164--173.
\newblock \doi{10.1371/journal.pcbi.0030015}.

\bibitem{Azevedoetal2006}
Azevedo, R. B.~R., Lohaus, R., Srinivasan, S., Dang, K.~K. \& Burch, C.~L.
  2006 Sexual reproduction selects for robustness and negative epistasis in
  artificial gene networks.
\newblock \emph{Nature} \textbf{440}, 87--90.
\newblock \doi{10.1038/nature04488}.

\bibitem{EspinosaSoto2010}
Espinosa-Soto, C. \& Wagner, A.  2010 Specialization can drive the evolution of
  modularity.
\newblock \emph{PLoS Computational Biology} \textbf{6}, e1000,719.
\newblock \doi{DOI 10.1371/journal.pcbi.1000719}.

\bibitem{JainKrug2007}
Jain, K. \& Krug, J.  2007 Deterministic and stochastic regimes of asexual
  evolution on rugged fitness landscapes.
\newblock \emph{Genetics} \textbf{175}, 1275--88.
\newblock \doi{10.1534/genetics.106.067165}.

\bibitem{Ostrowski2005}
Ostrowski, E.~A., Rozen, D.~E. \& Lenski, R.~E.  2005 Pleiotropic effects of
  beneficial mutations in \emph{{Escherichia} coli}.
\newblock \emph{Evolution} \textbf{59}, 2343--2352.
\newblock \doi{10.1111/j.0014-3820.2005.tb00944.x}.

\bibitem{Wagneretal2008}
Wagner, G., Kenney-Hunt, J., Pavlicev, M., Peck, J., Waxman, D. \& Cheverud, J.
   2008 Pleiotropic scaling of gene effects and the `cost of complexity'.
\newblock \emph{Nature} \textbf{452}, 470--472.
\newblock \doi{10.1038/nature06756}.

\bibitem{Skellettetal2005}
Skellett, B., Cairns, B., Geard, N., Tonkes, B. \& Wiles, J.  2005 Maximally
  rugged {NK} landscapes contain the highest peaks.
\newblock In H.-G. Beyer, ed., \emph{Proceeding GECCO '05 Proceedings of the
  2005 conference on Genetic and evolutionary computation}  pp. 579--584. New
  York, NY: Association for Computing Machinery.

\bibitem{FrankelYoung1998}
Frankel, A.~D. \& Young, J. A.~T.  1998 {HIV-1: {Fifteen} proteins and an RNA}.
\newblock \emph{Annual Review of Biochemistry} \textbf{67}, 1--25.

\bibitem{Hanetal2004}
Han, J. D.~J., Bertin, N., Hao, T., Goldberg, D.~S., Berriz, G.~F., Zhang,
  L.~V., Dupuy, D., Walhout, A. J.~M., Cusick, M.~E., Roth, F.~P. \emph{et~al.}
   2004 Evidence for dynamically organized modularity in the yeast
  protein-protein interaction network.
\newblock \emph{Nature} \textbf{430}, 88--93.
\newblock \doi{10.1038/nature02555}.

\bibitem{Doolittle2009}
Doolittle, R.~F., Jiang, Y. \& Nand, J.  2008 Genomic evidence for a simpler
  clotting scheme in jawless vertebrates.
\newblock \emph{Journal of Molecular Evolution} \textbf{66}, 185--196.
\newblock \doi{10.1007/s00239-008-9074-8}.

\bibitem{Andersonetal1981}
Anderson, S., Bankier, A., Barrell, B., Debruijn, M., Coulson, A., Drouin, J.,
  Eperon, I., Nierlich, D., Row, B., Sanger, F. \emph{et~al.}  {1981} Sequence
  and organization of the human mitochondrial genome.
\newblock \emph{Nature} \textbf{{290}}, {457--465}.
\newblock \doi{10.1038/290457a0}.

\bibitem{WagnerAltenberg1996}
Wagner, G.~P. \& Altenberg, L.  1996 Complex adaptations and the evolution of
  evolvability.
\newblock \emph{Evolution} \textbf{50}, 967--976.

\bibitem{Solowetal2000}
Solow, D., Burnetas, A., Tsai, M. \& Greenspan, N.~S.  2000 On the expected
  performance of systems with complex interactions among components.
\newblock \emph{Complex Systems} \textbf{12}, 423--456.
\newblock \doi{10.1007/978-3-540-74205-0\_31}.

\bibitem{Bonhoeffer2004}
Bonhoeffer, S., Chappey, C., Parkin, N.~T., Whitcomb, J.~M. \& Petropoulos,
  C.~J.  2004 Evidence for positive epistasis in {HIV}-1.
\newblock \emph{Science} \textbf{306}, 1547--1550.
\newblock \doi{10.1126/science.1101786}.

\bibitem{Mani2008}
Mani, R., {St. Onge}, R.~P., {Hartman IV}, J.~L., Giaever, G. \& Roth, F.~P.
  2008 Defining genetic interaction.
\newblock \emph{Proceedings of the National Academy of Sciences of the United
  States of America} \textbf{105}, 3461--3466.
\newblock \doi{10.1073/pnas.0712255105}.

\bibitem{Elena1997}
Elena, S.~F. \& Lenski, R.  1997 Test of synergistic interactions among
  deleterious mutations in bacteria.
\newblock \emph{Nature} \textbf{390}, 395--397.
\newblock \doi{10.1038/37108}.

\bibitem{Lenski1999}
Lenski, R.~E., Ofria, C., Collier, T.~C. \& Adami, C.  1999 Genome complexity,
  robustness and genetic interactions in digital organisms.
\newblock \emph{Nature} \textbf{400}, 661--664.
\newblock \doi{10.1038/23245}.

\bibitem{BurchChao2000}
Burch, C.~L. \& Chao, L.  2000 Evolvability of an {RNA} virus is determined by
  its mutational neighbourhood.
\newblock \emph{Nature} \textbf{406}, 625--628.
\newblock \doi{10.1038/35020564}.

\bibitem{Wichman1999}
Wichman, H.~A., Badgett, M.~R., Scott, L.~A., Boulianne, C.~M. \& Bull, J.~J.
  1999 Different trajectories of parallel evolution during viral adaptation.
\newblock \emph{Science} \textbf{285}, 422--424.
\newblock \doi{10.1126/science.285.5426.422}.

\bibitem{Eyre-walker2007}
Eyre-Walker, A. \& Keightley, P.~D.  2007 The distribution of fitness effects
  of new mutations.
\newblock \emph{Nature Reviews Genetics} \textbf{8}, 610--618.
\newblock \doi{10.1038/nrg2146}.

\bibitem{vanNimwegenetal1999}
van Nimwegen, E., Crutchfield, J.~P. \& Huynen, M.  1999 Neutral evolution of
  mutational robustness.
\newblock \emph{Proc Natl Acad Sci U S A} \textbf{96}, 9716--20.

\bibitem{Gillespie2004}
Gillespie, J.  2004 \emph{Population Genetics: A Concise Guide}.
\newblock Baltimore, MD: Johns Hopkins University Press.

\bibitem{Desaietal2007}
Desai, M.~M., Fisher, D.~S. \& Murray, A.~W.  2007 The speed of evolution and
  maintenance of variation in asexual populations.
\newblock \emph{Current Biology} \textbf{17}, 385--394.
\newblock \doi{10.1016/j.cub.2007.01.072}.

\bibitem{DesaiFisher2007}
Desai, M.~M. \& Fisher, D.~S.  2007 Beneficial mutation-selection balance and
  the effect of linkage on positive selection.
\newblock \emph{Genetics} \textbf{176}, 1759--1798.
\newblock \doi{DOI 10.1534/genetics.106.067678}.

\bibitem{Fogleetal2008}
Fogle, C.~A., Nagle, J.~L. \& Desai, M.~M.  2008 Clonal interference, multiple
  mutations and adaptation in large asexual populations.
\newblock \emph{Genetics} \textbf{180}, 2163--2173.
\newblock \doi{DOI 10.1534/genetics.108.090019}.

\bibitem{Barricketal2009}
Barrick, J.~E., Yu, D.~S., Yoon, S.~H., Jeong, H., Oh, T.~K., Schneider, D.,
  Lenski, R.~E. \& Kim, J.~F.  2009 Genome evolution and adaptation in a
  long-term experiment with escherichia coli.
\newblock \emph{Nature} \textbf{461}, 1243--U74.
\newblock \doi{10.1038/nature08480}.

\bibitem{Lenskietal1991}
Lenski, R., Rose, M., Simpson, S. \& Tadler, S.  1991 Long-term experimental
  evolution in \it {E}scherichia coli. \rm {I}. {A}daptation and divergence
  during 2,000 generations.
\newblock \emph{American Naturalist} \textbf{138}, 1315--1341.

\bibitem{ZeyldeVisser2001}
Zeyl, C. \& DeVisser, J.~A.  2001 Estimates of the rate and distribution of
  fitness effects of spontaneous mutation in saccharomyces cerevisiae.
\newblock \emph{Genetics} \textbf{157}, 53--61.

\bibitem{Weissman2009}
Weissman, D.~B., Desai, M.~M., Fisher, D.~S. \& Feldman, M.~W.  2009 The rate
  at which asexual populations cross fitness valleys.
\newblock \emph{Theoretical Population Biology} \textbf{75}, 286--300.
\newblock \doi{10.1016/j.tpb.2009.02.006}.

\bibitem{MacLean2009}
MacLean, R.~C. \& Buckling, A.  2009 The distribution of fitness effects of
  beneficial mutations in \emph{{Pseudomonas} aeruginosa}.
\newblock \emph{PLoS Genetics} \textbf{5}, e1000,406.
\newblock \doi{10.1371/journal.pgen.1000406}.

\bibitem{WilkeAdami2001}
Wilke, C.~O. \& Adami, C.  2001 Interaction between directional epistasis and
  average mutational effects.
\newblock \emph{Proceedings of the Royal Society of London B} \textbf{268},
  1469--1474.
\newblock \doi{10.1098/rspb.2001.1690}.

\bibitem{WilkeLenskiAdami2003}
Wilke, C.~O., Lenski, R.~E. \& Adami, C.  2003 Compensatory mutations cause
  excess of antagonistic epistasis in {RNA} secondary structure folding.
\newblock \emph{BMC Evolutionary Biology} \textbf{3}, 3.
\newblock \doi{10.1186/1471-2148-3-3}.

\bibitem{Bershteinetal2006}
Bershtein, S., Segal, M., Bekerman, R., Tokuriki, N. \& Tawfik, D.~S.  2006
  Robustness-epistasis link shapes the fitness landscape of a randomly drifting
  protein.
\newblock \emph{Nature} \textbf{444}, 929--932.
\newblock \doi{10.1038/nature05385}.

\bibitem{Beerenwinkeletal2007}
Beerenwinkel, N., Pachter, L., Sturmfels, B., Elena, S.~F. \& Lenski, R.~E.
  2007 Analysis of epistatic interactions and fitness landscapes using a new
  geometric approach.
\newblock \emph{{BMC} Evolutionary Biology} \textbf{7}.
\newblock \doi{10.1186/1471-2148-7-60}.

\bibitem{Sanjuanetal2006}
Sanjuan, R., Forment, J. \& Elena, S.~F.  2006 {In silico predicted robustness
  of viroid RNA secondary structures. II. Interaction between mutation pairs}.
\newblock \emph{Molecular Biology and Evolution} \textbf{23}, 2123--2130.
\newblock \doi{10.1093/molbev/msl083}.

\bibitem{segreetal2005}
Segre, D., DeLuna, A., Church, G.~M. \& Kishony, R.  2005 Modular epistasis in
  yeast metabolism.
\newblock \emph{Nature Genetics} \textbf{37}, 77--83.
\newblock \doi{doi:10.1038/ng1489}.

\bibitem{DonnellyWeber1985}
Donnelly, P. \& Weber, N.  1985 {The Wright-Fisher model with temporally
  varying selection and population size}.
\newblock \emph{Journal of Mathematical Biology} \textbf{22}, 21--29.
\newblock \doi{10.1007/BF00276544}.

\bibitem{Lindemannetal2002}
Lindemann, B.~F., Klug, C. \& Schwienhorst, A.  2002 Evolution of bacteriophage
  in continuous culture: a model system to test antiviral gene therapies for
  the emergence of phage escape mutants.
\newblock \emph{Journal of Virology} \textbf{76}, 5784--5792.
\newblock \doi{10.1128/JVI.76.11.5784-5792.2002}.

\end{thebibliography}










\newpage
\onecolumn
\pagestyle{empty}
\setcounter{figure}{0}
\renewcommand{\thefigure}{S\arabic{figure}}

\center {\Large Supplementary Material: Impact of Epistasis and Pleiotropy on Evolutionary Adaptation }

\noindent \center{\large \bf Supporting Figures}
\vskip 2cm

\begin{figure}[htbp]
\begin{center}
\includegraphics[width=0.5\textheight]{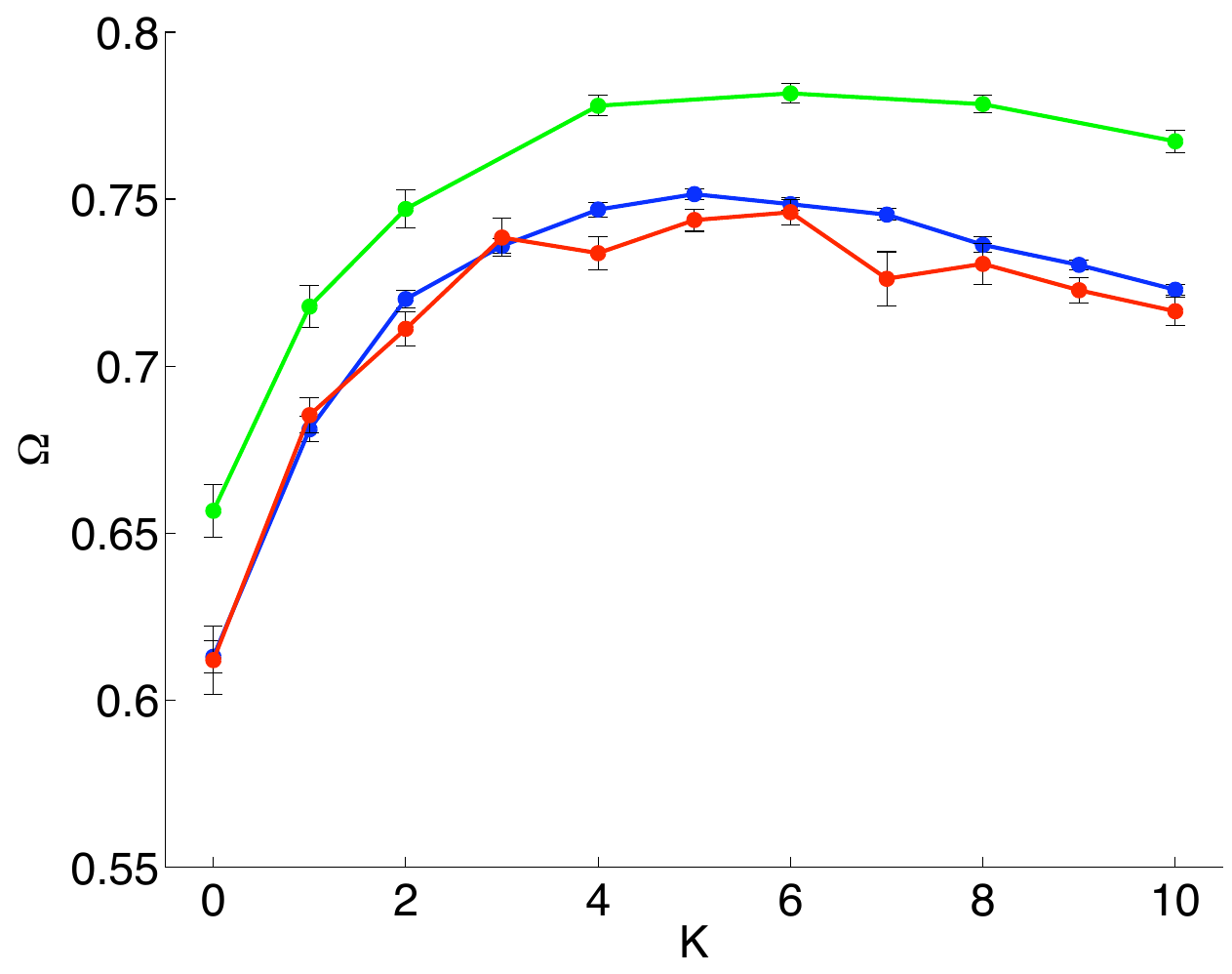}
\caption{\textbf{Comparison of attained fitness in three instances of the NK model.} 
Attained fitness as a function of $K$ in the traditional NK model where fitness is calculated as the arithmetic mean of the fitness components (green) compared to using the geometric mean (blue). Starting with a population of identical clones (red) rather than random genotypes of low fitness (blue) makes no significant difference.}
\label{supp_epi_mu}
\end{center}
\end{figure}

\begin{figure}[H]
\begin{center}
\includegraphics[width=0.75\textheight]{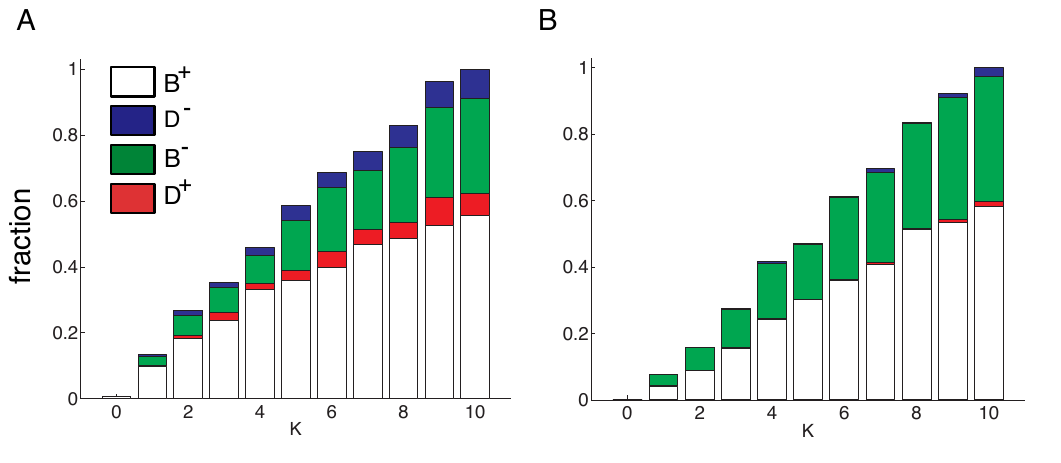}
\caption{\textbf{Fraction of types of the second substitution among all epistatic pairs.}
Height of bars shows the total fraction of epistatic substitutions. The distribution of types of the second substitution in epistatic pairs on the LOD changes when the mutation rate is changed. At the highest mutation rate a considerable fraction of epistatic substitutions are D$^+$ and D$^-$, while those fractions decrease at the lower mutation rates, leaving only B$^+$ and B$^-$ substitutions. Deleterious mutations are strongly selected against at the lowest mutation rate. B$^+$ (white), D$^+$ (red), B$^-$ (green), D$^-$ (blue). (A) $\mu=10^{-3}$. (B) $\mu=10^{-4}$.}
\label{second_mut}
\end{center}
\end{figure}

\begin{figure}[H]
\begin{center}
\includegraphics[width=0.75\textheight]{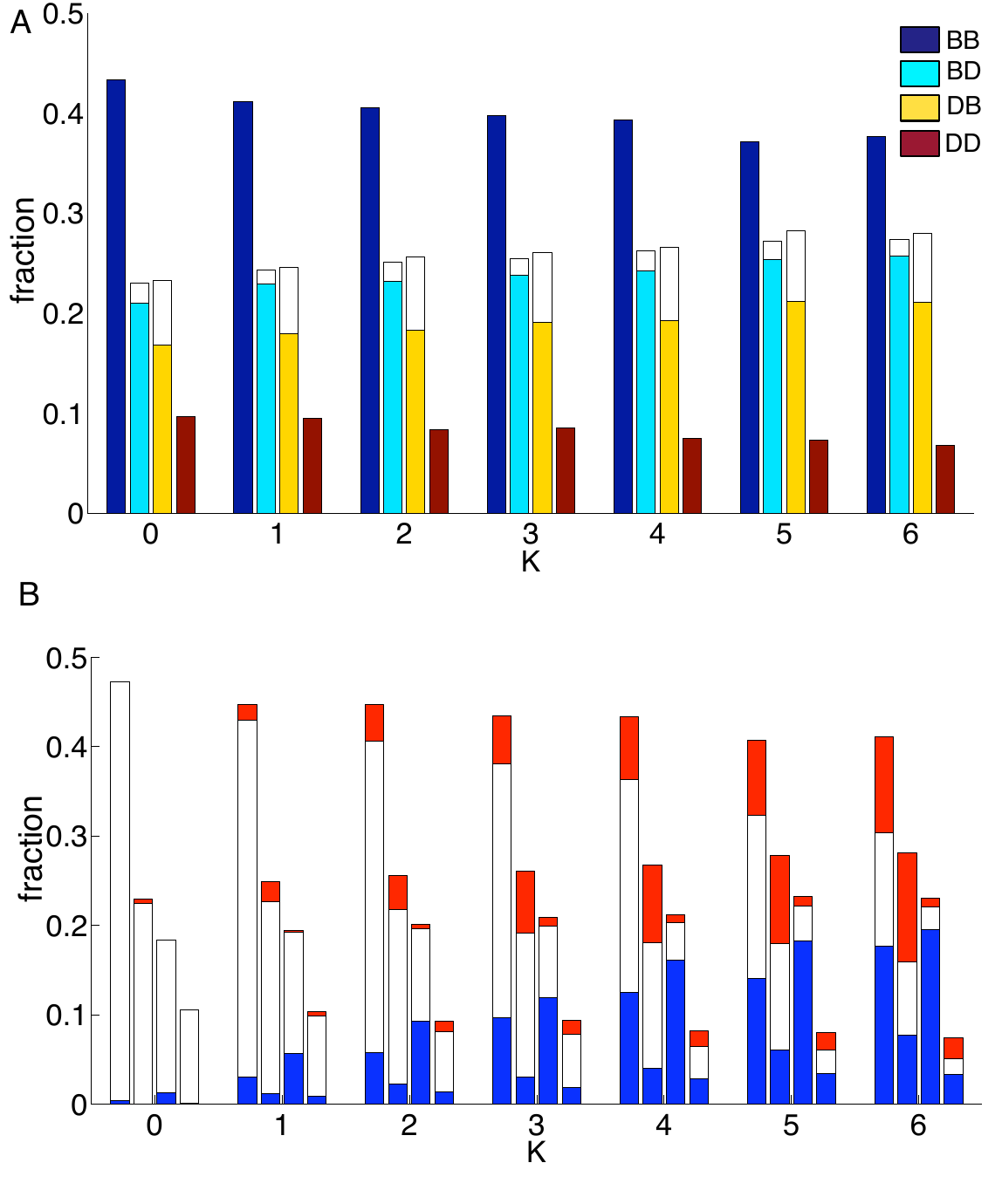}
\caption{\textbf{Distribution of pairs of substitutions.}
Distribution of BB, BD, DB, and DD pairs of substitutions on the LOD for $\mu=10^{-2}$.
(A): Fraction of pairs of substitutions of each type (indicated in the legend) vs. $K$. These fractions include non-interacting as well as epistatic pairs. The white part of the bars are due to reversal mutations only (reversal pairs). 
(B): Fraction of mutational pairs as in A, but with the negative (red), positive (blue), and zero epistasis between the pairs indicated. Positive epistasis dominate DB pairs, while both positive and negative epistasis is prominent among BB, BD, and DD pairs.} \label{pairs_types_distribution}
\end{center}
\end{figure}

\begin{figure}[H]
\begin{center}
\includegraphics[width=0.75\textheight]{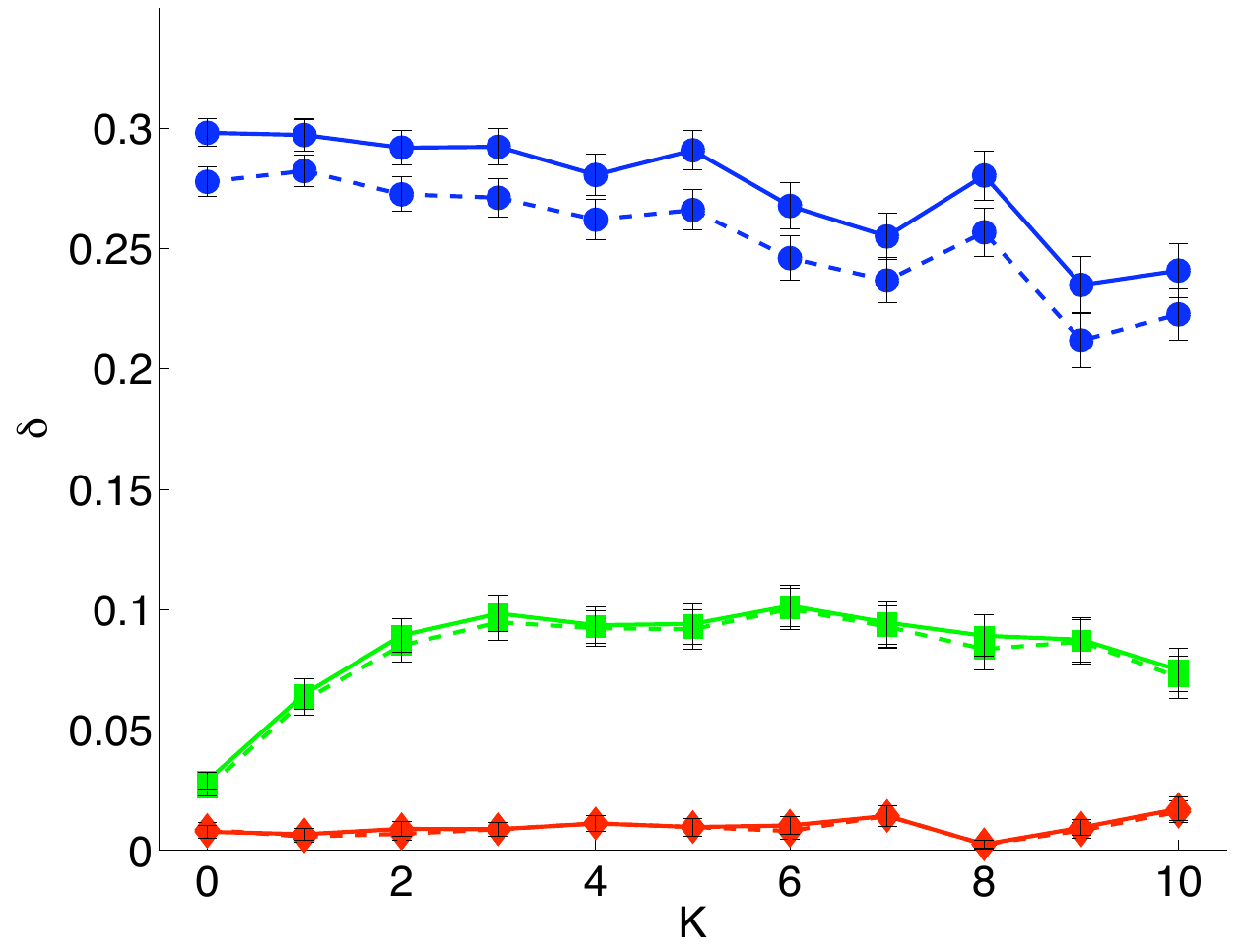}
\caption{\textbf{Fraction of deleterious substitutions.} 
Fraction of deleterious substitutions as a function of $K$ for different mutation rates. $\mu=10^{-2}$ (blue), $\mu=10^{-3}$ (green), and $\mu=10^{-4}$ (red). The solid lines show all deleterious substitutions including reversals mutations, while the dashed lines show deleterious substitutions excluding reversals. Population size is $5,000$, $N=20$, and the replacement rate is 10\%.}
\label{supp_epi_mu}
\end{center}
\end{figure}

\begin{figure}[H]
\begin{center}
\includegraphics[width=0.75\textheight]{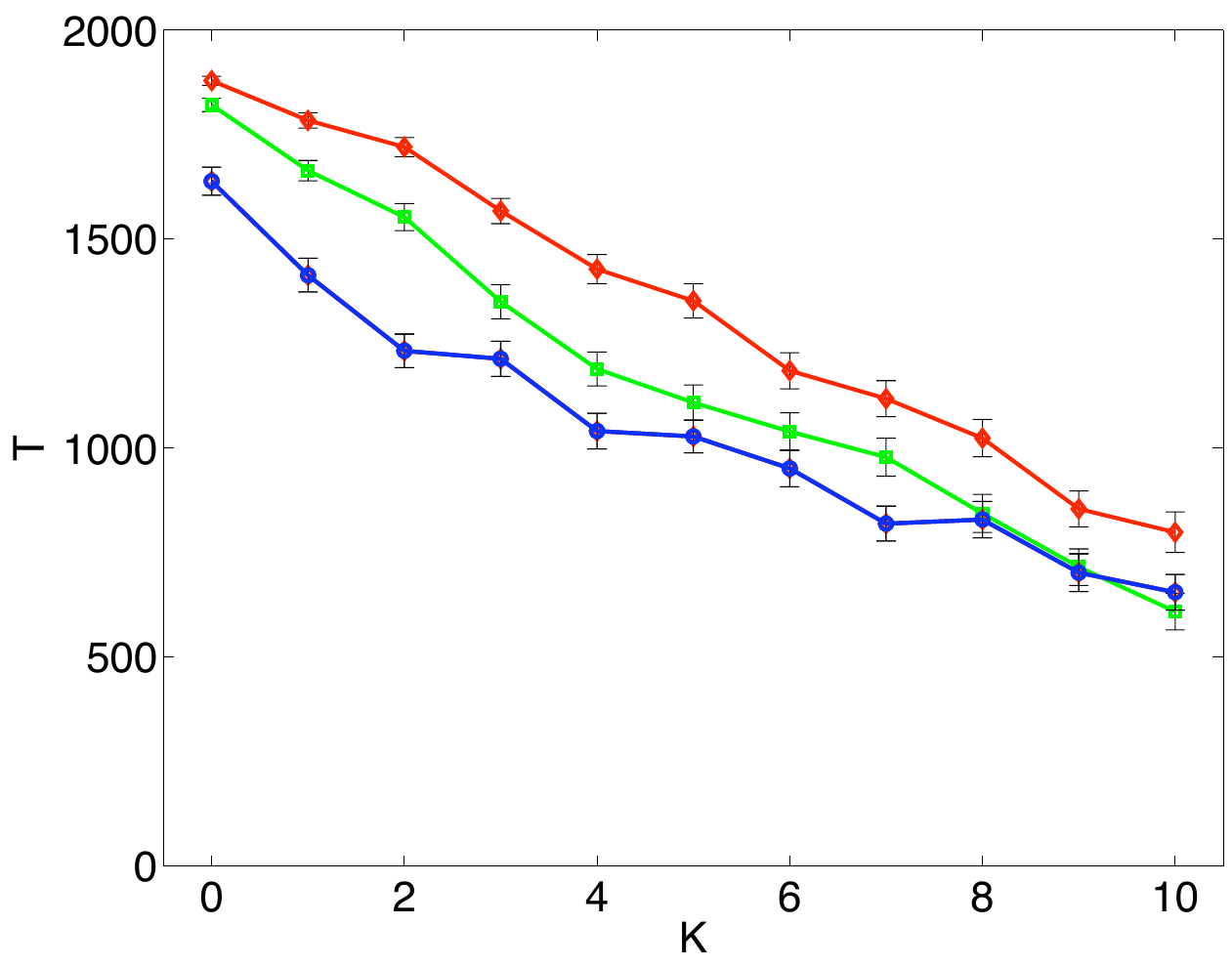}
\caption{\textbf{Average time to reach attained fitness.}
The time to reach the attained fitness (averaged over 200 replicate runs) is a decreasing function of $K$ ($\mu=10^{-2}$  (blue), $\mu=10^{-3}$ (green), and $\mu=10^{-4}$ (red).} \label{time_to_Omega}
\end{center}
\end{figure}

\begin{figure}[H]
\begin{center}
\includegraphics[width=0.5\textheight]{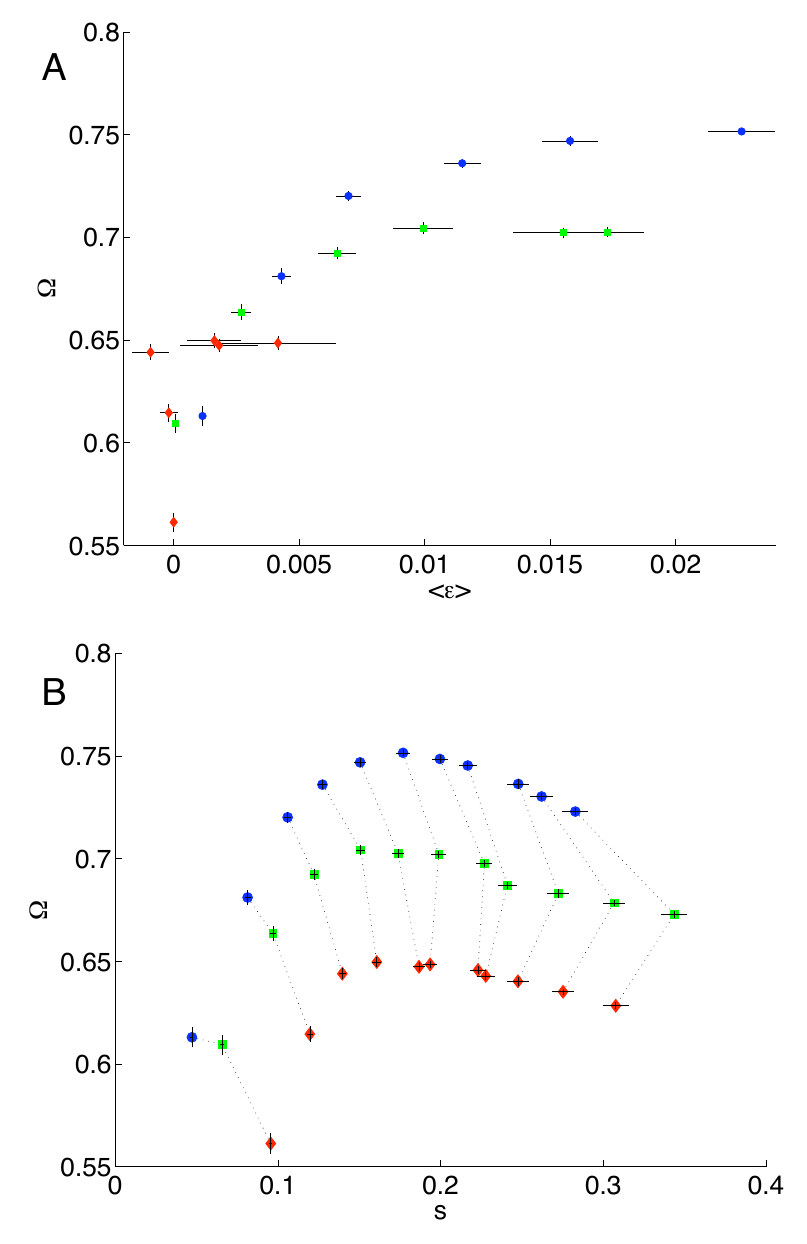}
\caption{\textbf{Comparison of how attained fitness $\Omega$ depends on biological observables.} A: Attained fitness as a function of mean epistasis on the LOD $\la\varepsilon\ra$ . For clarity, only $K\in [0,5]$ are included.  Colours for different mutation rates as in Fig. S4,  and error bars are standard error. B: Attained fitness $\Omega$ as a function of strength of selection $s$, for three different mutation rates (as in A). Points obtained with the same $K$ are joined by a dotted line. } \label{Omega_vs_s_and_e}
\end{center}
\end{figure}

\begin{figure}[H]
\includegraphics[width=0.7\textheight]{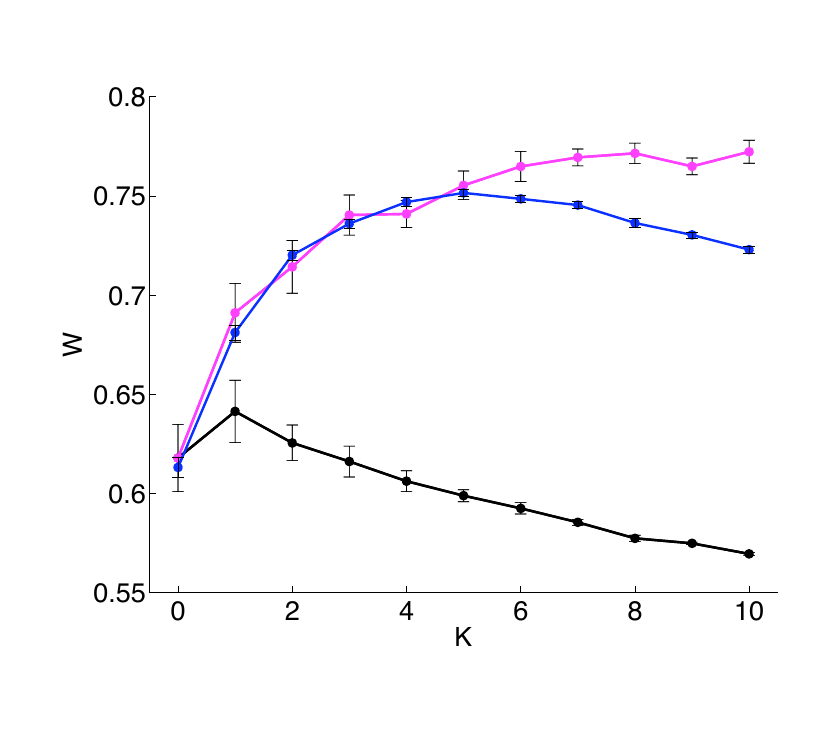}
\caption{\textbf{Mean and global peak height.}
For $K\leq5$ the attained fitness for $\mu=10^{-2}$ (blue) is equal to the maximum fitness of the landscape (magenta), indicating that when the mutation supply rate is high enough, the population is able to locate the global peak. At $K>5$ the population becomes stuck on other lower peaks. The mean fitness of peaks (black) is a decreasing function of $K$ for $K\geq1$, and has no bearing on the population's ability to adapt. Peaks are identified as those genotypes whose $N$ one-mutation neighbors have lower fitness.}
\label{peakheights}
\end{figure}

\begin{figure}[H]
\begin{center}
\includegraphics[width=0.5\textheight]{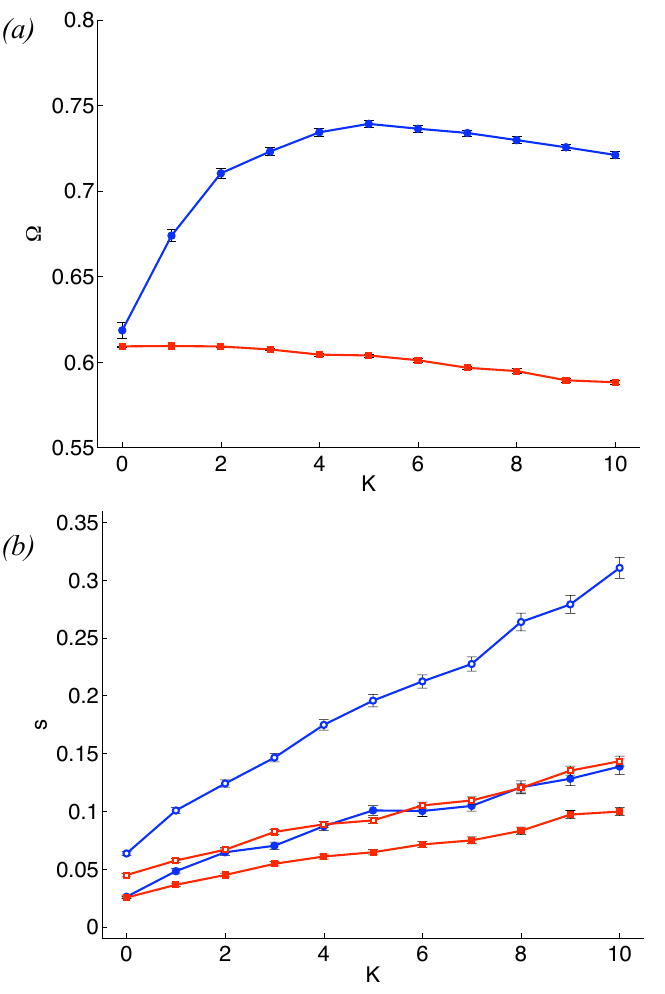}
\caption{\textbf{Attained fitness and strength of selection in the peak-normalized NK landscape.}
(A) Adaptation in a normalized NK landscape where the highest peak is normalized to the $K=0$ highest peak for all $K$. Blue: standard fitness landscape, red: normalized landscape. As $K$ increases, it is more and more difficult for the adaptive process to locate the peak in both landscapes, but the increase in peak height compensates for this effect up to moderate $K$. (B) Despite the reduced landscape amplitude (range in fitness values) that normalizing to $K=0$ produces, selection coefficients still increase approximately linearly with $K$ (red lines) compared to the standard landscape (blue lines), though at a shallower slope. This non-zero slope is a result of the higher frequency of peaks in the landscape, which makes the paths to the peaks steeper the higher $K$ is. Open symbols: beneficial mutations, solid symbols: deleterious mutations. All runs use $\mu=10^{-2}$.} \label{Omega_norm}
\end{center}
\end{figure}

\begin{figure}[H]
\begin{center}
\includegraphics[width=0.75\textheight]{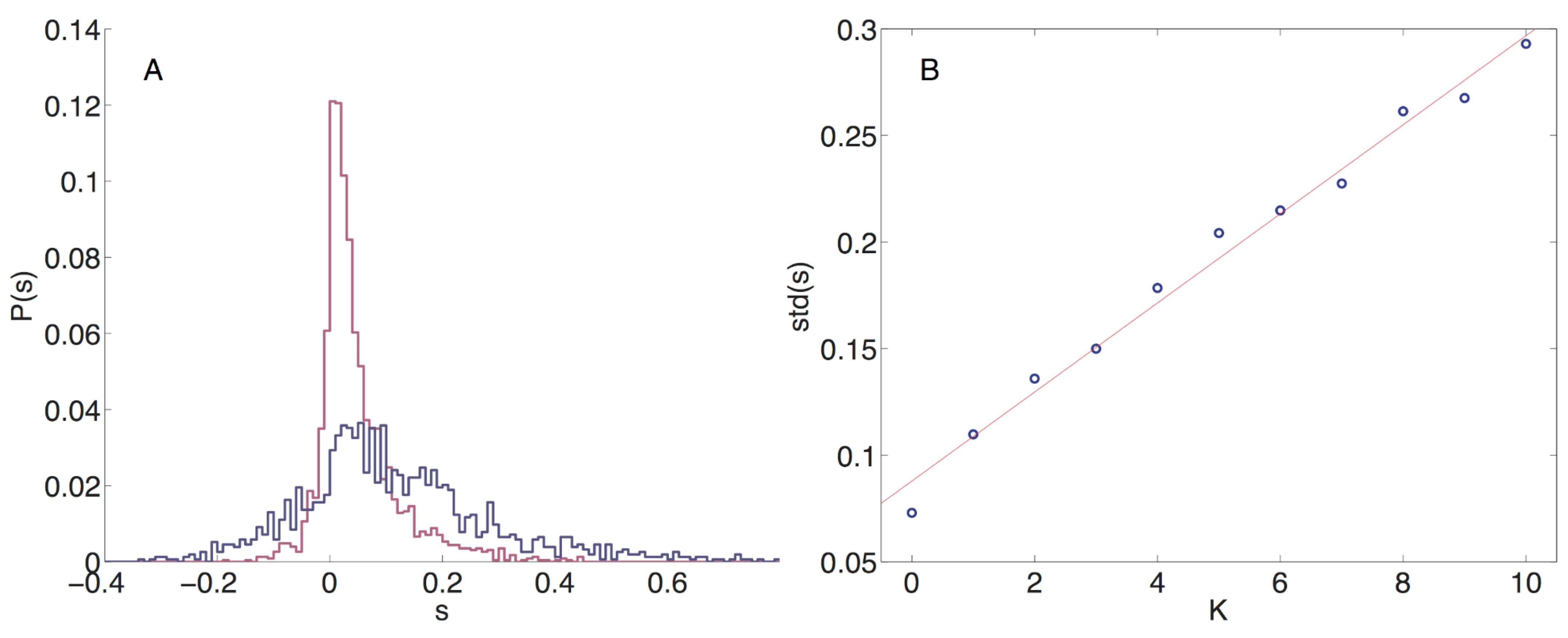}
\caption{\textbf{Distribution of selection coefficients.}
The larger $K$ is, the broader the distribution of the selection coefficients is. (A) Comparison of the distribution of selection coefficients for all 200 runs for two different values of $K$ reveals more deleterious substitutions of greater effect, $s<0$, for $K=4$ (blue) compared to $K=0$ (purple), but also that the effect of beneficial substitutions is increased even more at $K=4$. (B) The standard deviation of selection coefficients is a linear function of $K$. $\mu=10^{-2}$.}
\label{s_dist}
\end{center}
\end{figure}

\begin{figure}[H]
\begin{center}
\includegraphics[width=0.75\textheight]{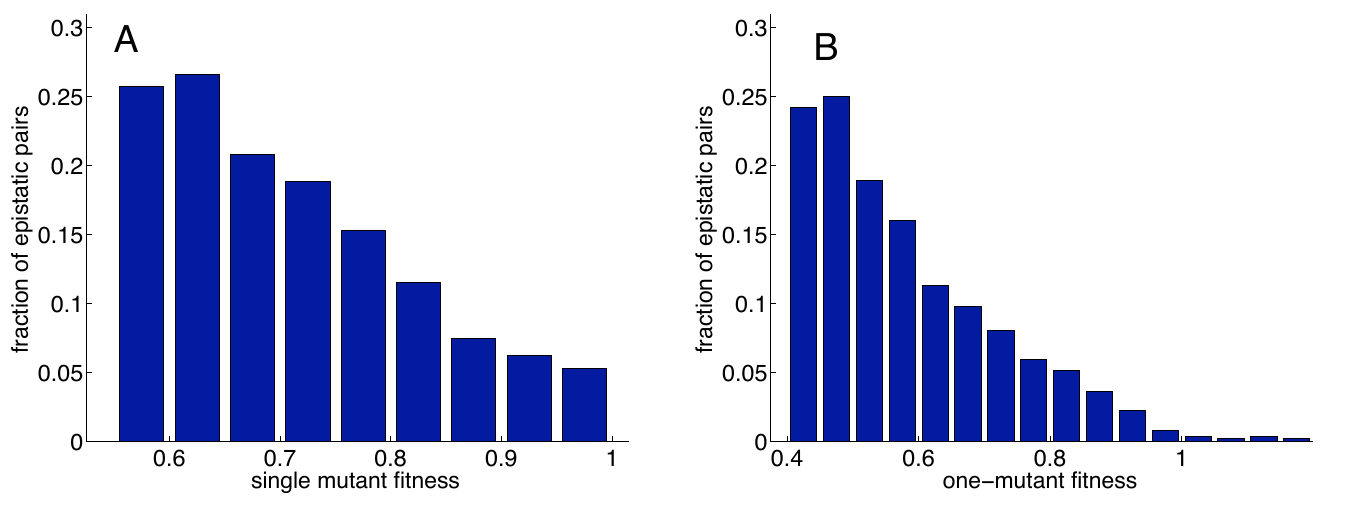}
\caption{\textbf{Fraction of epistatic pairs vs. deleterious effect}
(A) Fraction of pairs of mutations of high-fitness NK haplotypes with epistasis
$|\varepsilon|> 0.08$, for all 679 peaks in a landscape with $N=20$ and $K=4$, as a function of the single mutant fitness normalized to wild-type fitness (258,020 pairs). (B) Fraction of pairs of yeast gene knockouts with epistasis $|\varepsilon|> 0.08$ as a function of the single (normalized) mutant fitness. To obtain this figure, we used the original data from Costanzo et al. 2010 [30]  and measured epistasis using Eq.~[3] of the main text (5,481,706 pairs). In the latter study, a knockout could create an increased fitness for the cell, occasionally leading to a mutant fitness larger than the wild-type, while this was impossible in the NK landscape because we only used peak haplotypes.}
\label{frac-epi}
\end{center}
\end{figure}

\end{document}